\documentclass[twocolumn]{autart}  % Comment this line out if you need a4paper

\usepackage{graphicx}
\usepackage{epsfig} % for postscript graphics files
\usepackage{epstopdf}
\DeclareGraphicsExtensions{.xps}
\usepackage{scrextend}
\usepackage{amsmath}
\usepackage{amssymb}
\usepackage{float}
\usepackage{color}
\usepackage{caption}
\usepackage{subcaption}
\begin{document}

\newtheorem{Theo}{Theorem}
\newtheorem{Lem}{Lemma}
\newtheorem{Alg}{Algorithm}
\newtheorem{Assum}{Assumption}
\newtheorem{Rem}{Remark}
\newtheorem{Cor}{Corollary}
\newtheorem{Def}{Definition}
\newtheorem{Prob}{Problem}
\newtheorem{Pro}{Proposition}
\newtheorem{Exa}{Example}

\begin{frontmatter}

\title{On Decidability of Existence of Nonblocking Supervisors Resilient to Smart Sensor Attacks}

\thanks[footnoteinfo]{Corresponding author: Rong Su. Tel. +65 6790-6042. Fax: +65 6793-3318. The support from Singapore Ministry of Education Tier 1 Academic Research Grant 2018-T1-001-245 (RG 91/18) and from Singapore National Research Foundation Delta-NTU Corporate Lab Program (SMA-RP2) are gratefully acknowledged.}
\author[Paestum]{Rong Su}\ead{rsu@ntu.edu.sg}

\address[Paestum]{School of Electrical \& Electronic Engineering, Nanyang Technological University, 50 Nanyang Avenue, Singapore 639798.}  % Please supply                                              

 
\begin{abstract}                          % Abstract of not more than 200 words
Cybersecurity of discrete event systems (DES) has been gaining more and more attention recently, due to its high relevance to the so-called 4th industrial revolution that heavily relies on data communication among networked systems. One key challenge is how to ensure system resilience to sensor and/or actuator attacks, which may tamper data integrity and service availability. In this paper we focus on some key decidability issues related to smart sensor attacks.  We first present a sufficient and necessary condition that ensures the existence of a smart sensor attack, which reveals a novel  demand-supply relationship between an attacker and a controlled plant, represented as a set of risky pairs. Each risky pair consists of a damage string desired by the attacker and an observable sequence feasible in the supervisor such that the latter induces a sequence of control patterns, which allows the damage string to happen. It turns out that each risky pair can induce a smart weak sensor attack. Next, we show that, when the plant, supervisor and damage language  are regular,  it is possible to remove all such risky pairs from the plant behaviour, via a genuine encoding scheme, upon which we are able to establish our key result that the existence of a nonblocking supervisor resilient to smart sensor attacks is decidable. To the best of our knowledge, this is the first result of its kind in the DES literature on cyber attacks. The proposed decision process renders a specific synthesis procedure that guarantees to compute a resilient supervisor whenever it exists, which so far has not been achieved in the literature.            
\end{abstract}

\end{frontmatter}

%%%%%%%%%%%%%%%%%%%%%%%%%%%%%%%%%% Section   %%%%%%%%%%%%%%%%%%%%%%%%%%%%%%%%%%

\section{Introduction\label{GD}}
Cyber-attack resilience refers to properties of service availability and
data integrity. With the continuous advancement of information and
communications technology (ICT), in particular, the recent 5G-based
IoT technologies, we are enjoying unprecedented connectivity around
the world. Nevertheless, the threat of cyber attacks that may potentially
cause significant damage to human lives and properties has more
frequently become the center of attention, and has been attracting lots
of research from different communities. Basically, an attacker aims to
inflict damage on a target system by disrupting its control loop. This
could be achieved either by intercepting and changing the controller’s
input signals (in terms of sensor attacks), or by intercepting and changing
the controller’s output signals (in terms of actuator attacks), or by completely blocking the data transmission between the controller
and the plant (in terms of denial-of-service attacks). An attack can be
either brute-force, e.g., via hardware destruction or signal jamming,
or covert (or stealthy), i.e., to inflict damage without being detected by relevant monitoring mechanisms. 

A good survey of cyber attacks and cyber defence with a systems-and-control perspective can be found in \cite{Dibaji2019}. Typically, linear systems are considered in existing works that rely on system identification and control
techniques. Within the DES community, most works rely on the control
system setup introduced in the Ramadge-Wonham supervisory control
paradigm \cite{RW87}, where the plant generates
observable outputs, received by the supervisor via an observation channel,
and each control command specified as a set of allowed (or disallowed)  events
is generated by the supervisor and fed to the plant via a command
channel. The plant nondeterministically picks one event from a received
control command and execute{\color{black}s} it. The event execution process is assumed to be asynchronous, i.e., up to one event execution at each time instant, and instantaneous.  Unlike attacks in time-driven systems
described in \cite{Dibaji2019}, attacks under consideration in a DES
aim to change the {\color{black}sequence} of events in specific system runs. 

There are two
different streams of research on cyber attacks and resilient control. The
first stream refers to a set of black-box methods that treat attacks as undesirable (either intentional or unintentional) uncontrollable and mostly unobservable disturbances to a given closed-loop system. Existing works include, e.g., a game theoretical approach \cite{RMT14}, fault-tolerance based approaches such as \cite{Gao2019} and \cite{Wakaiki2019} on sensor attacks, \cite{Carvalho2016} on actuator attacks, and \cite{Lima2017} \cite{Carvalho2018} \cite{Lima2018} on sensor+actuator attacks, and transducer-based modelling and synthesis approaches such as \cite{WangandPajic2019a} \cite{WangandPajic2019b}. In the black-box methods, system vulnerability is typically modelled by concepts similar to diagnosability described in, e.g., \cite{Sam95}, and system resilience bears similarity to fault tolerant control described in, e.g., \cite{PSL11} \cite{ABCCM14}, that concerns whether there is a supervisor that can perform satisfactorily under the worst case attack scenarios.  
The second stream refers to a set of white-box methods, aiming to develop a specific ``smart'' attack
model that ensures certain intuitive properties such as covertness and
guaranteed (strong or weak) damage infliction. Existing works include, e.g., \cite{Su2017} \cite{Su2018} \cite{Goes2017} \cite{Goes2019} \cite{Goes2020} {\color{black}\cite{Goes2020}} \cite{FritzandZhang2018} on smart sensor attacks,  \cite{Lin2019a} \cite{Zhu2019} and \cite{Lin2020} on smart actuator attacks, and \cite{Lin2019b} \cite{Khoumsi2019} and \cite{LinandSu2020} on smart sensor+actuator attacks. With such smart attack models, existing research works address the impact of a specific attack on the closed-loop behaviour, the vulnerability of  a system to such an attack, and finally the resilience of a supervisor to a concerned attack.

After examining those existing works on smart cyber attacks, it is clear that most works focus on how to derive a proper smart attack model. Various synthesis algorithms have been proposed under relevant assumptions. Nevertheless, the existence of a supervisor that is resilient to {\color{black}all possible} smart cyber attacks is still open for research. {\color{black}In \cite{GML19} {\color{black}\cite{GLM21}} the authors present synthesis methods for resilient control against a specific sensor attack model described by a finite-state automaton in different scenarios. Thus, the synthesized supervisor is not designed to be resilient to all possible smart sensor attacks. In case of a worst-case sensor attack scenario, no smartness in terms of, e.g., covertness, is considered by the authors.} There are a few heuristic synthesis approaches proposed in the literature, e.g., \cite{Su2018} proposes one algorithm against smart  sensor attacks, \cite{Lin2019b} proposes one algorithm that generates a resilient supervisor whose state set is bounded by a known value, and \cite{Zhu2019} presents an algorithm to synthesize a supervisor, which is control equivalent to an original supervisor and resilient to smart actuator attacks. But none of those existing algorithms can guarantee to find one resilient supervisor, if it exists. That is, when those algorithms terminate and return an empty solution, it does not necessarily mean that there is no solution.

Before any attempt of overcoming a complexity challenge in order to derive a resilient solution, it is critical to answer a computability question first, that is, how to decide whether a solution  exists.  To address this important decidability issue, in this paper we focus only on sensor attacks, but hoping that our derived result may shed light on research of other types of attacks. We follow a sensor attack model proposed in \cite{Su2018}, which associates each observed sequence from the plant $G$ with an altered observable sequence that becomes the input of a given supervisor. After slightly improving the concept of {\em attackability} originally introduced in \cite{Su2018} and the corresponding definition of smart sensor attacks, our first contribution is to identify conditions that ensure the existence of a smart sensor attack. It turns out that the existence of a smart weak sensor attack, which is not necessarily regular (i.e., representable by a finite-state automaton), is solely determined by the existence of at least one risky pair that consists of a damage string desired by the attacker and an observable sequence feasible in the supervisor such that the latter induces a sequence of control patterns, which allows the concerned damage string to happen. Because any strong sensor attack is also a weak attack, the existence of such a risky pair becomes the sufficient and necessary condition for the existence of a smart sensor attack. In \cite{Su2017} and its journal version \cite{Su2018}, by imposing language normality to the closed-loop behaviour, it is shown that the supremal smart sensor attack language can be synthesi{\color{black}zed}, whenever it exists, upon which a specific smart sensor attack model can be derived. In \cite{Goes2017} and its journal version \cite{Goes2020}, the language normality is dropped, and it is shown that a smart sensor attack model (not necessarily supremal) can be synthesized via a special insertion-deletion attack structure, whenever it exists. However, none of these works reveals the aforementioned demand-supply relationship reflected in risky pairs that capture the nature of sensor attacks. Due to this insightful concept of risky pairs, our second contribution is to show that the existence of a nonblocking controllable and observable supervisor that is resilient to {\color{black}all regular} smart sensor attacks is decidable. To this end, we develop a genuine encoding mechanism that reveals all possible sequences of control patterns required by a regular sensor attack and all sequences of control patterns feasible in the plant, allowing us to remove the set of all risky pairs from the plant behaviour. After that, we introduce a language-based concept of {\em nonblocking resilient supervisor candidate} and its automaton-counterpart {\em control feasible sub-automaton} that does not contain any risky pair, upon which we are able to decide the existence of a resilient supervisor. As our third contribution, the proposed decision process renders a concrete synthesis procedure that guarantees to compute a nonblocking supervisor resilient to smart sensor attacks, whenever it exists.

The remainder of the paper is organized as follows. In Section 2 {\color{black}we present a motivation example. Then in Section 3} we review the basic concepts and operations of discrete event systems introduced in \cite{Won07}, followed by a specific smart sensor attack model, where the concept of {\em attackability} is introduced. Then we present a sufficient and necessary condition to ensure a smart sensor attack in Section {\color{black}4}, where the key concept of risky pairs is introduced. After that, we present a sufficient and necessary condition for the existence of a nonblocking supervisor that is resilient to smart sensor attacks in Section {\color{black}5}, and show that this sufficient and necessary condition is verifiable in Section {\color{black}6}, which finally establishes the decidability result for the existence of a resilient supervisor.  Conclusions are drawn in Section {\color{black}7}. {\color{black}Long proofs for Theorems 1-4 are shown in the Appendix.}  

{\color{black}
\section{A motivation example - attack of navigator}
Image that Bob would like to ride his autonomous car from his home to his office. There is a GPS navigator installed inside his car. The navigator first generates a shortest path based on traffic conditions, then guides the vehicle to make required turns at planned junctions. However, Bob's friend Peter plans to play a prank on Bob by tricking the navigator to lead Bob's car to another location via GPS spoofing, shown in Figure 
\ref{fig:Cyber-Security-00001}. 
\begin{figure}[htb]
    \begin{center}
      \includegraphics[width=0.3\textwidth]{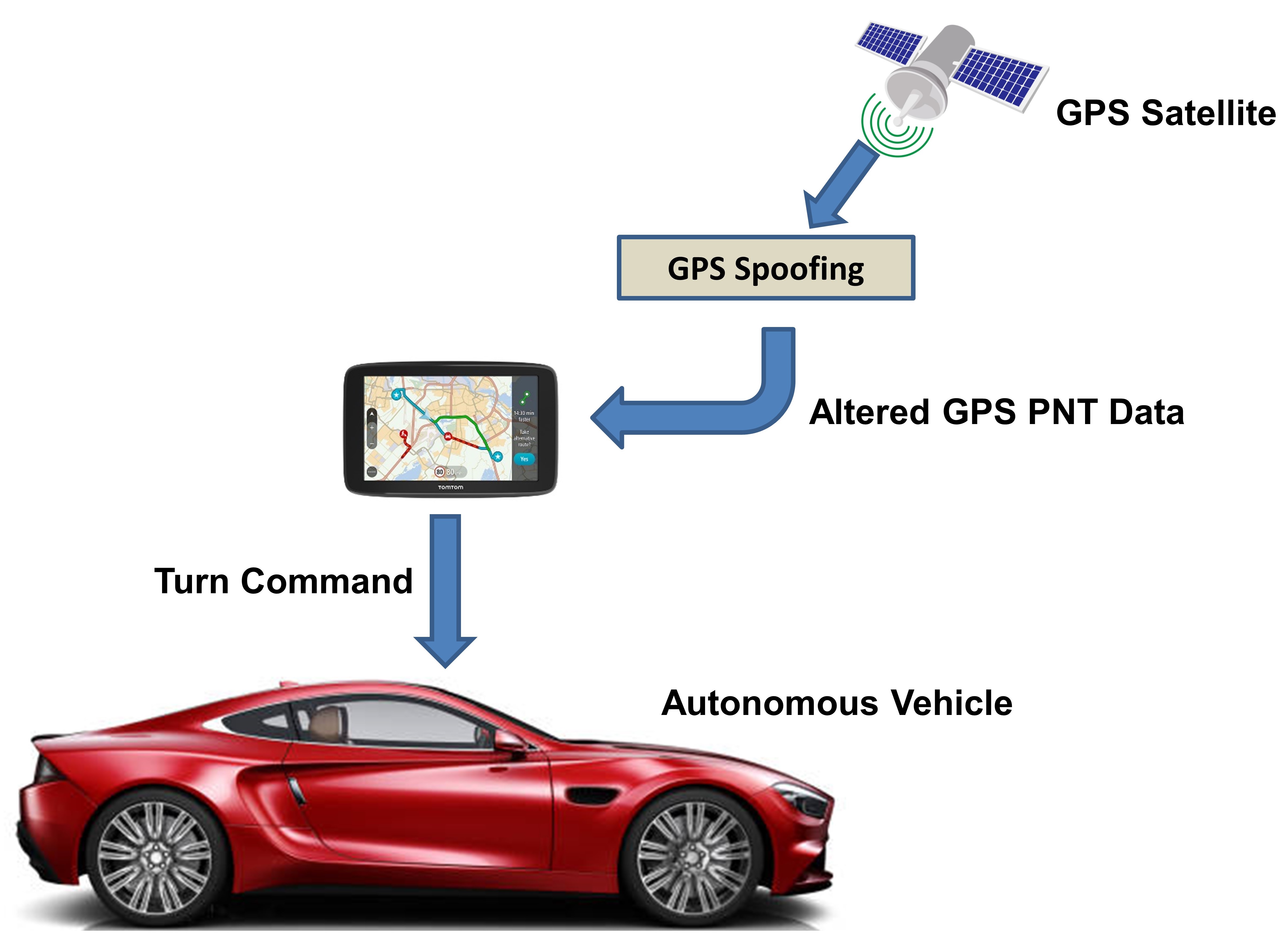}
    \end{center}
    \caption{Example 0: An example of attack of navigator}
    \label{fig:Cyber-Security-00001}
\end{figure}
Peter has the city road map and also knows Bob's home address and office address. In addition, he has a navigator of the same model as the one installed in Bob's car. Thus, by running his own navigator over the same origin-destination pair, Peter will know Bob's route plan.  Figure \ref{fig:Cyber-Security-00002} depicts the system setup,    
\begin{figure}[htb]
    \begin{center}
      \includegraphics[width=0.4\textwidth]{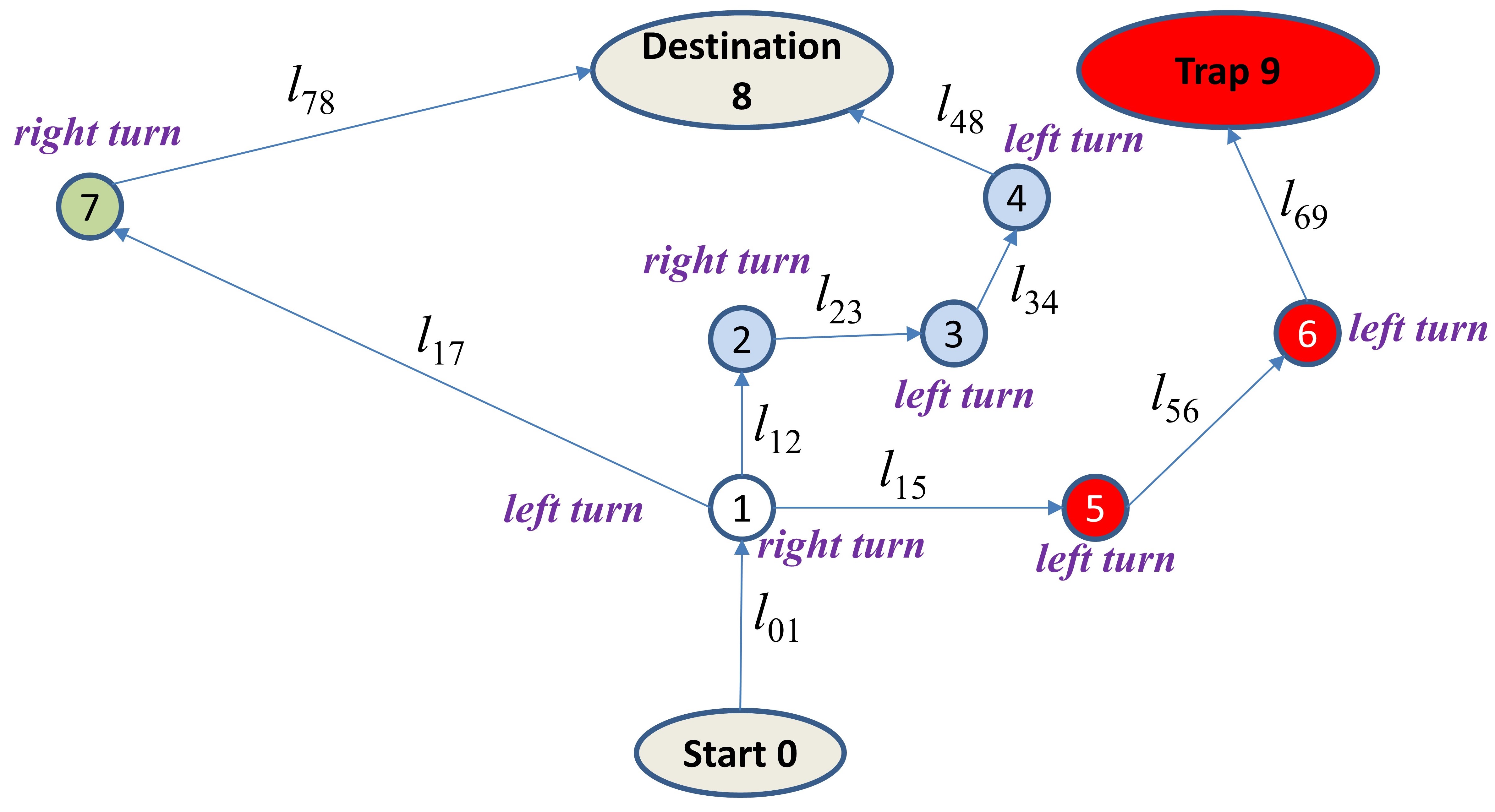}
    \end{center}
    \caption{Example 0: Road network setup}
    \label{fig:Cyber-Security-00002}
\end{figure}
where Bob's home is at ``start 0'' node, his office is at ``Destination 8'' node, and the prank location is at ``Trap 9'' node. Each symbol ``$l_{ij}$'' denotes the length of the road segment between node $i$ and node $j$. In addition, the position of each node in the map is publicly known. By simply running the navigator of the same model, Peter knows that Bob's car will choose the shortest path ``$0\rightarrow 1\rightarrow 2\rightarrow 3\rightarrow 4\rightarrow 8$'', which leads to the following navigation commands: (1) {\em right turn} (at node 2), followed by (2) {\em left turn} (at node 3), followed by (3) {\em left turn} (at node 4). To trick the navigator to issue the same sequence of commands but at incorrect junctions, say,  {\em right turn} at node 1, followed by {\em left turn} at node 5, and followed by {\em left turn} at node 6, Peter only needs to buy a GPS spoofing device available in the market that can send fake GPS position signals to Bob's navigator. Peter can easily determine the spoofed GPS position signal as follows. Assume that $p_0$ denotes the position of ``start 0'' node, and $p(t)\in\mathbb{R}^2$ and $p^a(t)\in\mathbb{R}^2$ denote the actual and spoofed GPS positions in a 2D map. The travel distance made by Bob's car between time $t_0$ and time $t$ ($t\geq t_0$)  can be calculated by a simple line integral shown below:
\[\int_{t_0}^t\frac{p(\tau)}{||p(\tau)||}\cdot dp(\tau),\]
where $p(t)$ is the parametric function of the path from node ``start 0'' to node 1,   $p(t_0)=p_0$, $||p(t)||$ denotes the magnitude of $p(t)$ and `$\cdot$' is the dot product of vectors. For each $p(t)$, Peter uniquely determines the value of $p^a(t)$, which is the parametric function of the path from node ``start 0'' to node 2  with $p^a(t_0)=p_0$, such that  
\[\frac{\int_{t_0}^t\frac{p^a(\tau)}{||p^a(\tau)||}\cdot dp^a(\tau)}{\int_{t_0}^t\frac{p(\tau)}{||p(\tau)||}\cdot dp(\tau)}=\frac{l_{01}+l_{12}}{l_{01}},\]
which ensures that, when the navigator receives the spoofed signal, indicating that node 2 is reached, the actual node reached is node 1. With a similar GPS position spoofing scheme, Peter can misguide Bob's car to reach node 9, instead of node 8, without being detected. Such GPS spoofing is one specific example of a {\em smart sensor attack}, whose formal definition will be given in the next section. Intuitively, it contains the following basic characteristics: by knowing sufficient information in advance, an attacker can trick a victim to issue the correct order of commands but at incorrect states (or locations), which leads to an unwanted consequence.    

If Bob somehow knows that Peter will use GPS spoofing to play a trick on his car, he can simply choose the path ``$0\rightarrow 1\rightarrow 7\rightarrow 8$''. In this case, even Peter knows this new path, he cannot spoof the GPS signals to trick Bob's car to node 9 without being detected, as the new path generates a new sequence of navigation commands: (1) {\em left turn} (at node 1), then (2) {\em right turn} (at node 7), which, no matter how Peter changes GPS position signals, cannot bring Bob to node 9. Such a path plan is a specific example of a {\em resilient supervisor} against smart sensor attacks, whose definition will be given later in this paper. Intuitively, a resilient supervisor will ensure that, for {\em\textbf{any}} sensor attack, either it can be detected before inflicting damage, or it will not lead to any damage. 

One big question is, for an arbitrary network, see, e.g., a road map of a small region in Singapore shown in Figure  \ref{fig:Cyber-Security-00003},
\begin{figure}[htb]
    \begin{center}
      \includegraphics[width=0.3\textwidth]{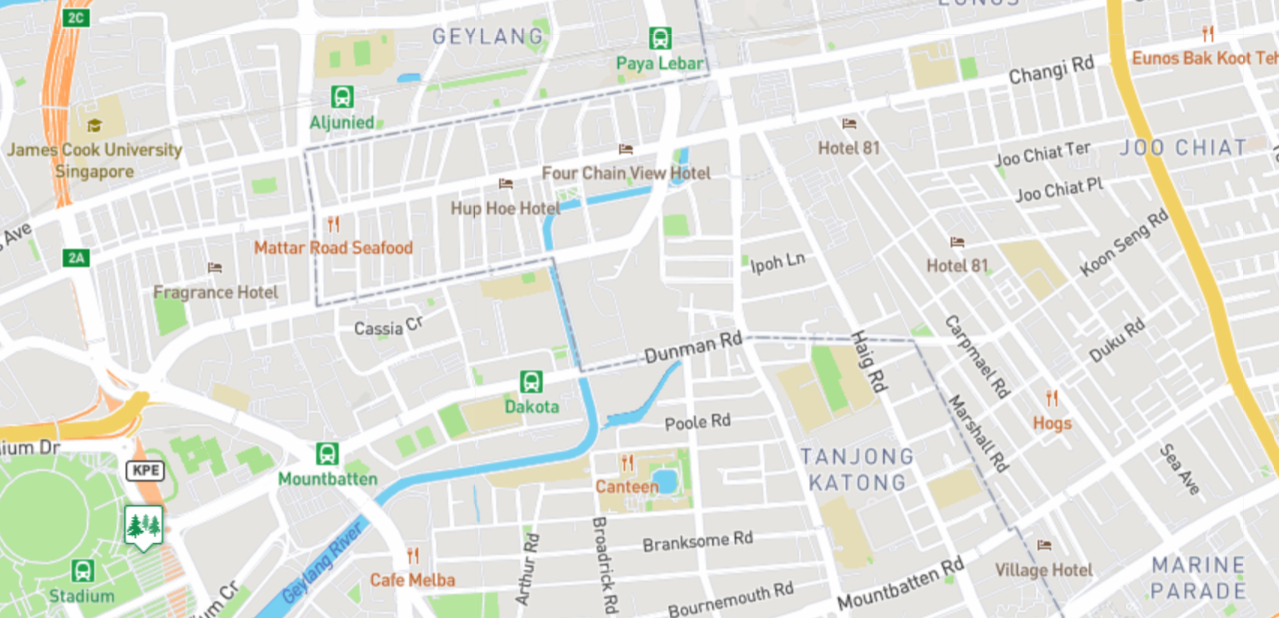}
    \end{center}
    \caption{Example 0: A more realistic road network}
    \label{fig:Cyber-Security-00003}
\end{figure}
how to decide whether a resilient path plan (or navigation supervisor) exists. In this paper, we will investigate this decidability problem against smart sensor attacks. The computational efficiency, i.e., the complexity issue, however, will not be addressed here.     
}

\section{A smart sensor attack model}
In this section we first recall basic concepts used in the Ramadge-Wonham supervisory control paradigm \cite{RW87}. Then we recall a smart sensor attack model introduced in \cite{Su2018}. Most notations in this paper follow \cite{Won07}.

\subsection{Preliminaries on supervisory control}
Given a finite alphabet $\Sigma$, let $\Sigma^*$ be the free monoid over $\Sigma$ with the empty string $\epsilon$ being the unit element and the string concatenation being the monoid binary operation. We use $\Sigma^+$ to denote non-empty strings in $\Sigma^*$, i.e., $\Sigma^+=\Sigma^*-\{\epsilon\}$. Given two strings $s,t\in\Sigma^*$, we say $s$
is a \emph{prefix substring} of $t$, written as $s\leq t$, if
there exists $u\in\Sigma^*$ such that $su=t$, where $su$ denotes
the concatenation of $s$ and $u$.  For any string $s'\in\Sigma^*$ with $s'\leq s$, we use $s/s'$ to denote the post substring $u\in\Sigma^*$ such that $s=s'u$. We use $|s|$ to denote the length of $s$, and by convention, $|\epsilon|=0$.  Any subset $L\subseteq\Sigma^*$ is called a \emph{language}.  The \emph{prefix closure} of
$L$ is defined as $\overline{L}=\{s\in\Sigma^*|(\exists t\in L)\,
s\leq t\}\subseteq\Sigma^*$. For each string $s\in \overline{L}$, let $En_{\overline{L}}(s):=\{\sigma\in\Sigma|s\sigma\in \overline{L}\}$ be the set of events that can extend $s$ in $\overline{L}$. Given
two languages $L,L'\subseteq\Sigma^*$, let
$LL':=\{ss'\in\Sigma^*|s\in L\,\wedge\, s'\in L'\}$ denote the
concatenation of two sets.  Let $\Sigma'\subseteq\Sigma$. A mapping $P:\Sigma^*\rightarrow\Sigma'^*$ is called the \emph{natural projection} with respect to $(\Sigma,\Sigma')$,
if
\begin{enumerate}
\item $P(\epsilon)=\epsilon$, 
\item $(\forall \sigma\in\Sigma)\, P(\sigma):=\left\{\begin{array}{ll} \sigma & \textrm{ if $\sigma\in\Sigma'$,}\\
    \epsilon & \textrm{ otherwise,}\end{array}\right.$ \item $(\forall s\sigma\in\Sigma^*)\, P(s\sigma)=P(s)P(\sigma)$.
\end{enumerate}
Given a language $L\subseteq\Sigma^*$, $P(L):=\{P(s)\in\Sigma'^*|s\in L\}$. The inverse image mapping of $P$ is
\[P^{-1}:2^{\Sigma'^*}\rightarrow 2^{\Sigma^*}:L\mapsto
P^{-1}(L):=\{s\in\Sigma^*|P(s)\in L\}.\] 
%Given $L_1\subseteq\Sigma_1^*$ and $L_2\subseteq\Sigma_2^*$, the \emph{synchronous product} of $L_1$ and $L_2$
%is defined as $L_1||L_2:=P_1^{-1}(L_1)\cap P_2^{-1}(L_2)$,
%where $P_1:(\Sigma_1\cup\Sigma_2)^*\rightarrow\Sigma_1^*$ and $P_2:(\Sigma_1\cup\Sigma_2)^*\rightarrow\Sigma_2^*$ are natural projections. Clearly,
%$||$ is commutative and associative.\\

A given target plant is modelled as a \emph{deterministic finite-state automaton}, $G=(X,\Sigma,\xi,x_0,X_m)$, where $X$ stands for the state set, $\Sigma$ for the
alphabet, $\xi:X\times\Sigma\rightarrow X$ for the (partial) transition function, $x_0$ for the initial state and $X_m\subseteq X$ for the
marker state set. We follow the notation in \cite{Won07}, and use $\xi(x,\sigma)!$ to denote that the transition $\xi(x,\sigma)$ is defined. For each state $x\in X$, let $En_G(x):=\{\sigma\in\Sigma|\xi(x,\sigma)!\}$ be the set of events enabled at $x$ in $G$.  The domain of $\xi$ can be extended to $X\times\Sigma^*$, where $\xi(x,\epsilon)=x$ for all $x\in X$, and $\xi(x,s\sigma):=\xi(\xi(x,s),\sigma)$. The \emph{closed} behaviour of $G$ is defined as $L(G):=\{s\in\Sigma^*|\xi(x_0,s)!\}$, and the \emph{marked} behaviour of $G$ is $L_m(G):=\{s\in L(G)|\xi(x_0,s)\in X_m\}$. $G$ is \emph{nonblocking} if $\overline{L_m(G)}=L(G)$. We will use $\mathbb{N}$ to denote natural numbers, $|X|$ {\color{black}(or $|G|$)} for the size of the state set $X$, and $|\Sigma|$ for the size of $\Sigma$. Given two finite-state automata $G_i=(X_i,\Sigma,\xi_i,x_{i,0},X_{i,m})$ ($i=1,2$), the \emph{meet} of $G_1$ and $G_2$, denoted as $G_1\wedge G_2$, is a (reachable) finite-state automaton whose alphabet is $\Sigma$ such that $L(G_1\wedge G_2)=L(G_1)\cap L(G_2)$ and $L_m(G_1\wedge G_2)=L_m(G_1)\cap L_m(G_2)$. A {\em sub-automaton} of $G$ is an automaton $G_{sub}=(X,\Sigma,\xi_{sub},x_0,X_m)$ such that 
\[(\forall x,x'\in X)(\forall \sigma\in\Sigma)\, \xi_{sub}(x,\sigma)=x'\Rightarrow \xi(x,\sigma)=x',\] 
that is, each transition of $G_{sub}$ must be a transition in $G$, but the opposite may not be true. When the transition map is $\xi:X\times\Sigma\rightarrow 2^X$, where $2^X$ denotes the power set of $X$, we call $G$ a {\em nondeterministic} finite-state automaton. If for each $x\in X$, there exists $s\in\Sigma^*$ such that $\xi(x,s)\cap X_m\neq\varnothing$, then $G$ is {\em co-reachable}. For the remainder of this paper, unless explicitly mentioned, all automata are assumed to be deterministic.

We now recall the concept of supervisors. Let
$\Sigma=\Sigma_c\dot{\cup}\Sigma_{uc}=\Sigma_o\dot{\cup}\Sigma_{uo}$, where $\Sigma_c$ ($\Sigma_o$) and $\Sigma_{uc}$ ($\Sigma_{uo}$) {\color{black}are disjoint, denoting} respectively
the sets of \emph{controllable} (\emph{observable}) and
\emph{uncontrollable} (\emph{unobservable}) events. For notational simplicity, let $\Sigma_o^{\epsilon}:=\Sigma_o\cup\{\epsilon\}$.  Let $\Gamma=\{\gamma\subseteq\Sigma|\Sigma_{uc}\subseteq\gamma\}$, where each $\gamma\in\Gamma$ is one {\em control pattern} (or {\em control command}). A supervisory control map of $G$ under partial observation $P_o:\Sigma^*\rightarrow\Sigma_o^*$ is defined as $V:P_o(L(G))\rightarrow \Gamma$. Clearly, \[(\forall s\in L(G))(\forall \sigma\in\Sigma_{uc})\, s\sigma\in L(G)\Rightarrow \sigma\in V(P_o(s)),\]
namely the supervisory control map $V$ never tries to disable an uncontrollable transition. In addition, 
\[(\forall s,s'\in L(G))\, P_o(s)=P_o(s')\Rightarrow V(P_o(s))=V(P_o(s')),\]
namely any two strings in $L(G)$ that are observably identical, their induced control patterns are equal. 

Let $V/G$ denote the closed-loop system of $G$ under the supervision of $V$, i.e.,
\begin{itemize}
\item $\epsilon\in L(V/G)$,
\item For all $s\in L(V/G)$ and $\sigma\in\Sigma$ \[s\sigma\in L(V/G)\iff s\sigma\in L(G)\wedge \sigma\in V(P_o(s)),\]
\item $L_m(V/G):=L_m(G)\cap L(V/G)$.
\end{itemize}
The control map $V$ is \emph{finitely representable} if $V$ can be described by a finite-state automaton, say $S=(Z,\Sigma,\delta,z_o,Z_m=Z)$, such that
\begin{itemize}
\item $L(S\wedge G)=L(V/G)$ and $L_m(S\wedge G)=L_m(V/G)$,
\item $(\forall s\in L(S))\, En_S(s):=\{\sigma\in\Sigma|s\sigma\in L(S)\}=V(s)$,
%\item $(\forall s\sigma\in L(S)\Sigma_{uo}\cap L(V/G))\, \delta(z_0,s)=\delta(z_0,s\sigma)$,
\item $(\forall s,s'\in L(S))\, P_o(s)=P_o(s')\Rightarrow {\color{black}\delta(z_0,s)=\delta(z_0,s')}$.
\end{itemize} 
The last condition indicates that $V(s)=En_S(s)=En_S(s')=V(s')$ if $P_o(s)=P_o(s')$. Such a supervisor $S$ can be computed by existing synthesis tools such as TCT \cite{TCT} or SuSyNA \cite{SuSyNA}. It has been shown  that, as long as a closed-loop language $K\subseteq L_m(G)$ is \emph{controllable} \cite{RW87}, \emph{observable} \cite{LW88} and $L_m(G)$-closed, i.e., $K=\overline{K}\cap L_m(G)$, there always exists a finitely-representable supervisory control map $V$ such that $L_m(V/G)=K$ and $L(V/G)=\overline{K}$. From now on we make the following assumption.
\begin{Assum} $V$ is nonblocking, i.e., $L(V/G)=\overline{L_m(V/G)}$, and finitely representable by $S$.\hfill $\Box$\end{Assum} 
We will use $V$ or $S$ interchangeably, depending on the context. They will be called a (nonblocking) {\em supervisor}.

\subsection{A smart sensor attack model}
We assume that an attacker can observe each observable event generated by the plant $G$, and replace the observable event with a sequence of observable events from $\Sigma_o^*$, including the empty string $\epsilon$, in order to ``fool'' the given supervisor $V$, known to the attacker.  Considering that in practice any event occurrence takes a non-negligible amount of time, it is impossible for an attacker to insert an arbitrarily long observable sequence to replace a received observable event. Thus, we assume that there exists a known number $n\in\mathbb{N}$ such that the length of any ``reasonable'' observable sequence that the attacker can insert is no more than $n$. Let $\Delta_n{\color{black}\subseteq}\{s\in\Sigma_o^*||s|\leq n\}$ be the set of all $n$-bounded observable sequences {\color{black} possibly inserted by an attacker}. A sensor attack is a total map $A:P_o(L(G))\rightarrow \Delta_n^*$, where
\begin{itemize}
\item $A(\epsilon)=\epsilon$,
\item {\color{black}$(\forall \sigma\in\Sigma_0)\, A(\sigma)\in\Delta_n$,}
\item ${\color{black}(\forall s\in P_o(L(G)))(\forall \sigma\in\Sigma_o)\, A(s\sigma)=A(s)A(\sigma)}$.
\end{itemize}
The first condition states that, before any observation is obtained, the attack cannot generate any non-empty output, because, otherwise, such a fake observation sequence may reveal the existence of an attack, if the plant has not started yet, whose starting time is unknown to the attacker.  The second {\color{black}and third} condition{\color{black}s} state that each received observation $\sigma\in\Sigma_o$ will trigger a fake string in $\Delta_n$. This model captures moves of insertion, deletion and replacement introduced in, e.g., \cite{Goes2017} \cite{Carvalho2018} \cite{Goes2020}. 

An attack model $A$ is {\em regular} if there exists a finite-state transducer $\mathcal{A}=(Y,\Sigma_o^{\epsilon}\times \Delta_n,\eta,I,O,y_0,Y_m)$, where  
$Y_m=Y$, $\eta:Y\times \Sigma_o^{\epsilon}\times\Delta_n\rightarrow Y$ is the (partial) transition mapping such that if $\eta(y,\sigma,u)!$ and $\sigma=\epsilon$ then $u=\epsilon$, i.e., if there is no observation input, then there should be no observation output. The functions $I:(\Sigma_o^{\epsilon}\times\Delta_n)^*\rightarrow \Sigma_o^*$ and $O:(\Sigma_o^{\epsilon}\times\Delta_n)^*\rightarrow \Delta_n^*$ are the \emph{input} and \emph{output} mappings, respectively, such that for each $\mu=(a_1,b_1)(a_2,b_2)\cdots (a_l,b_l)\in (\Sigma_o^{\epsilon}\times\Delta_n)^*$,  $I(\mu)=a_1 a_2\cdots a_l$ and $O(\mu)=b_1b_2\cdots b_l$. We require that, for each $\mu\in L(\mathcal{A})$, we have $A(I(\mu))=O(\mu)$ and $I(L(\mathcal{A}))= P_o(L(G))$. Since $A$ is a function, we know that for all $\mu,\mu'\in L(\mathcal{A})$, if $I(\mu)=I(\mu')$, then $O(\mu)=A(I(\mu))=A(I(\mu'))=O(\mu')$, that is, the same input should result in the same output. Notice that, in \cite{Su2017} \cite{Su2018}, an attack model is directly introduced as a finite-state transducer, which may not necessarily be representable by an attack map $A$, because a finite-transducer model allows nondeterminism, i.e., for the same observation input, an attacker may choose different attack moves, as long as they are allowed by the transducer model. In this sense, the attack model concerned in this paper is a special case of the one introduced in \cite{Su2017} \cite{Su2018}, and bears more resemblance to the model introduced in \cite{Goes2020}, as both treat an attack as a function. But since there exists a nondeterministic attack model if and only if there exists a deterministic one, the decidability results derived in this paper shall be applicable to nondeterministic attack models introduced in \cite{Su2017} \cite{Su2018}. {\color{black}To see this fact, it is clear that each deterministic model is a nondeterministic model. Thus, we only need to show that from each nondeterministic model we can derive at least one deterministic model. We will use a simple example  to illustrate the construction procedure. Assume that the nondeterministic attack model adopted from \cite{Su2018} is shown in Figure \ref{fig:Cyber-Security-10000}, which is a transducer. We first start from damage states (i.e., marker states), and perform co-reachability search to find all states in the nondeterministic model that satisfy the following two conditions: (1) each state is reachable from the initial state, (2) at each state, each observable event is associated with only one transition (denoting an attack move). After that, we perform reachability search upon those states derived from the first step and add new necessary states in so that the following condition holds: at each state, each observable event is associated with only one transition (denoting an attack move) if and only if it is associated with at least one transition in the original nondeterministic model. This construction will result in a deterministic smart sensor attack model, shown in Figure \ref{fig:Cyber-Security-10000}.  
\begin{figure}[htb]
    \begin{center}
      \includegraphics[width=0.4\textwidth]{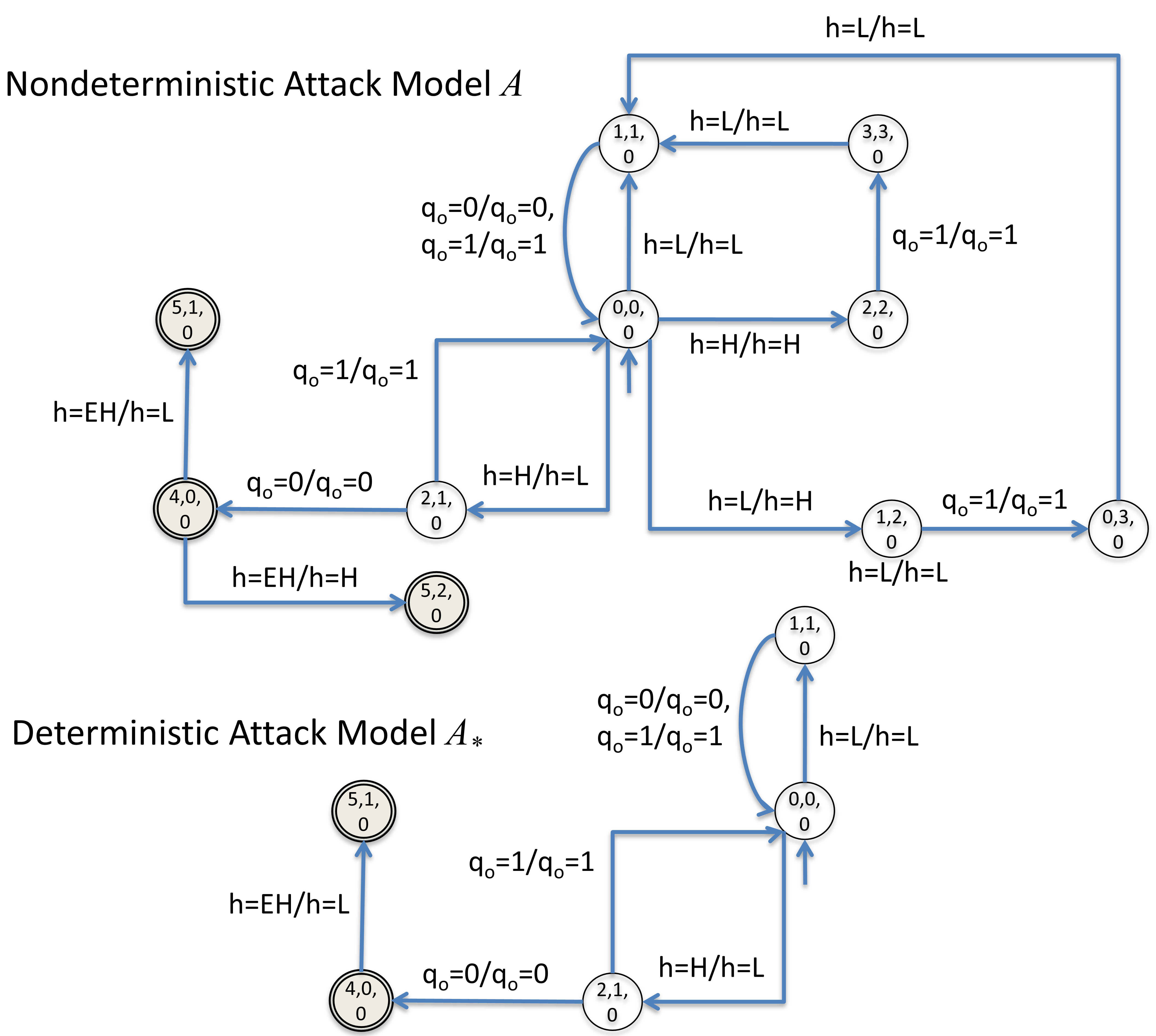}
    \end{center}
    \caption{Transforming a nondeterministic attack to a deterministic attack}
    \label{fig:Cyber-Security-10000}
\end{figure}
In this example, we can see that there can be several deterministic attack models derivable from the nondeterministic model.}      

\begin{Assum}Only regular attacks are considered.\end{Assum}

\begin{figure}[htb]
    \begin{center}
      \includegraphics[width=0.4\textwidth]{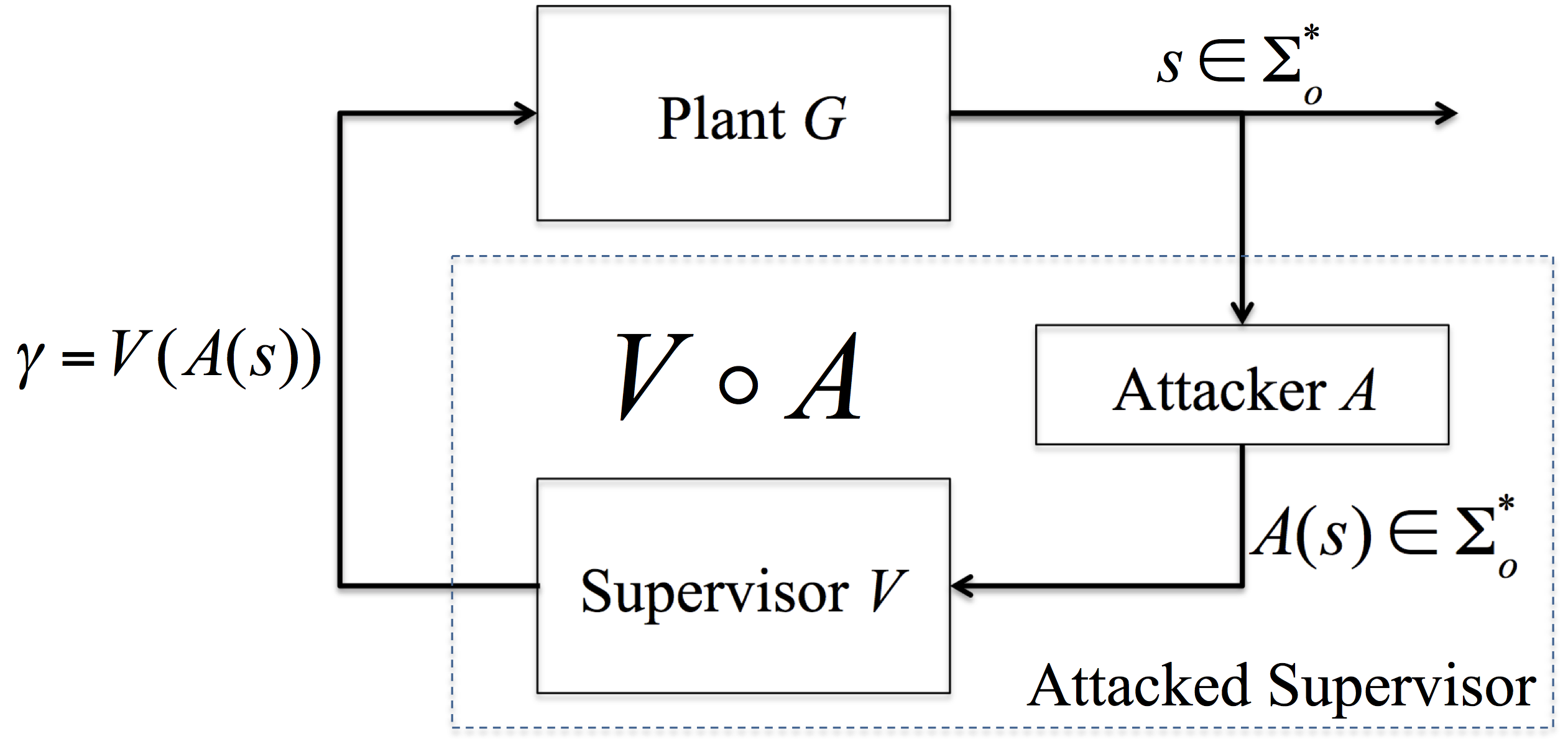}
    \end{center}
    \caption{The block diagram of a plant under attack}
    \label{fig:Cyber-Security-0}
\end{figure}

The combination of the attack $A$ and the supervisor $V$ forms a new supervisor $V\circ A:P_o(L(G))\rightarrow\Gamma$, where 
\[(\forall s\in P_o(L(G)))\, V\circ A(s):=V(A(s)).\]
We call $V\circ A$ an {\em attacked supervisor under $A$}. {\color{black}This definition consumes the standard style of one command per each received (fake) observation used in, e.g., \cite{Carvalho2018} \cite{GML19}, as a special case, when $n$ is set to 1.} The closed and marked behaviours, $L(V\circ A/G)$ and $L_m(V\circ A/G)$, of the closed-loop system $V\circ A/G$ are defined accordingly. We call $L(V\circ A/G)$ the {\em attacked language} of $V/G$ under $A$.
The closed-loop system is depicted in Figure \ref{fig:Cyber-Security-0}.

\begin{Def}\label{Def20}
\textnormal{Given a plant $G$ and a supervisor $V$ realized by $S$, let $L_{dam}\subseteq L(G)-L(V/G)$ be a damage language. The closed-loop system $V/G$ is {\em attackable} with respect to $L_{dam}$, if there exists an attack $A$, called a {\em smart sensor attack} of $V/G$, such that the following hold:
\begin{enumerate}
\item \textbf{Covertness}: Event changes made by $A$ are covert to the supervisor $V$, i.e., \begin{equation}A(P_o(L(V\circ A/G)))\subseteq L(S).\end{equation}
%and $A$ does not stop the plant's output, i.e., for all $s\in L(V\circ A/G)$ and $\sigma\in V(A(P_o(s)))\cap\Sigma_o$,
%\begin{equation}s\sigma\in L(G)\Rightarrow P_o(s)\sigma\in I(L(A)).\end{equation}
\item \textbf{Damage infliction}: $A$ causes ``damage'' to $G$, i.e.,
\begin{itemize}
\item strong attack: Any string may lead to damage:\begin{equation}L(V\circ A/G)=\overline{L(V\circ A/G)\cap L_{dam}};\end{equation}
\item weak attack: Some string may lead to damage: \begin{equation}L(V\circ A/G)\cap L_{dam}\neq\varnothing.\end{equation}
\end{itemize}
%\item \textbf{Control Feasibility}: The closed-loop language $L(V\circ A/G)$ is observable with respect to $(L(G), P_o)$ \cite{LW88}, i.e., for all $s,s'\in L(V\circ A/G)$ and $\sigma\in\Sigma$, if $s\sigma\in L(V\circ A/G)$, $P_o(s)=P_o(s')$ and $s'\sigma\in L(G)$, then $s'\sigma\in L(V\circ A/G)$.
\end{enumerate}
If $V/G$ is not attackable with respect to $L_{dam}$, then $V$ is {\em resilient} to smart attacks with respect to $L_{dam}$. \hfill $\Box$} 
\end{Def}

The concept of attackability introduced in Def. \ref{Def20} simplifies the concept of attackability introduced in \cite{Su2018} by, firstly, dropping the requirement of control existence, as $V\circ A$ automatically allows all uncontrollable transitions, thus, ensuring controllability, and secondly, {\color{black}dropping the normality requirement, as we directly model an attack as a function over the plant's observable behaviours, instead of a language used in \cite{Su2018} (which is equivalent to a nondeterministic attack), making the closed-loop language $L(V\circ A/G)$ observable automatically}. 

{\color{black}Remark: In \cite{Su2018}, a special subset of observable events called {\em protected events} is introduced, which is denoted by $\Sigma_{o,p}\subseteq\Sigma_o$, representing observable events in the plant that cannot be changed by any sensor attack. This feature makes the modeling framework more general. However, it diminishes the chance of having a smart sensor attack, due to the challenge of ensuring the covertness property, when the system trajectory $s\in L(V\circ A/G)$ is outside the legal language $L(V/G)$ and there are a few protected system output events that will inevitably  reveal the attack. Due to this complication, we lack a simple sufficient and necessary condition to characterize the existence of a smart sensor attack, making the subsequent study of the existence of a supervisor resilient to such smart sensor attacks infeasible. To overcome this challenge, we could restrict the damage language $L_{dam}$ to be $L(G)-L(V/G)$, i.e., any string outside $L(V/G)$ is a damage string. This will allow us to relax the covertness property to be \[A(P_o(L(V\circ A/G)\cap L(V/G)))\subseteq L(S),\] that is, an attacker does not need to make any event change, after the system trajectory is outside $L(V/G)$, as damage has been inflicted. Then all results presented in this paper will still be valid. So a user of this theory has two options for the system setup, that is, either $\Sigma_{o,p}=\varnothing$ and $L_{dam}\subseteq L(G)-L(V/G)$, as adopted in Def. \ref{Def20} of this paper, or $\Sigma_{o,p}\neq\varnothing$ and $L_{dam}=L(G)-L(V/G)$. }

Let $\mathcal{F}(G,V, L_{dam})$ be the collection of all attacked languages caused by smart sensor attacks. Clearly, $(\mathcal{F}(G,V, L_{dam}),\subseteq)$ is a partially ordered set.
%and by Zorn's lemma, we know that it has at least one maximal element, which is called a {\em maximal attacked language}.  
When  {\color{black}$L(V\circ A/G)$ is required to be {\em normal} \cite{LW88}, i.e., only observable events may be disabled}, and {\color{black}the} attack model $A$ is nondeterministic, i.e., for the same observable input, $A$ may have more than one output choice, it has been shown in \cite{Su2018} that $(\mathcal{F}(G,V, L_{dam}),\subseteq)$ {\color{black}over all smart strong attacks} becomes an upper semilattice, and the supremal strong attacked language $L(V\circ A_*/G)$ exists such that for any smart {\color{black}strong} sensor attack $A$, we have $L(V\circ A/G)\subseteq L(V\circ A_*/G)$. In this case, the supremal strong attacked language is computable, as shown in \cite{Su2018}. {\color{black}With a similar spirit, the supremal  weak attacked language exists and is also computable, as briefly mentioned in the conclusion} of \cite{Su2018}.  {\color{black}When only deterministic attack models are adopted, it turns out that the supremal deterministic attack model may not always exist. However, by computing the supremal nondeterministic smart weak attacked language first, which induces a finite-state transducer, as shown in \cite{Su2018}, we can show that a deterministic attack model derivable from the finite-state transducer by applying the transformation procedure shown in Figure \ref{fig:Cyber-Security-10000} results in a maximal attacked language in $\mathcal{F}(G,V, L_{dam})$. Detailed arguments are omitted here, due to limited space and the focus of this paper that is not about supremality or maximality of attacked languages. In the example depicted in Figure \ref{fig:Cyber-Security-10000}, the illustrated deterministic smart attack model results in a maximal attacked language. However, there is no supremal attacked language induced by a deterministic smart weak attack.}

\section{A sufficient and necessary condition for the existence of a smart sensor attack}
\begin{figure}[htb]
    \begin{center}
      \includegraphics[width=0.45\textwidth]{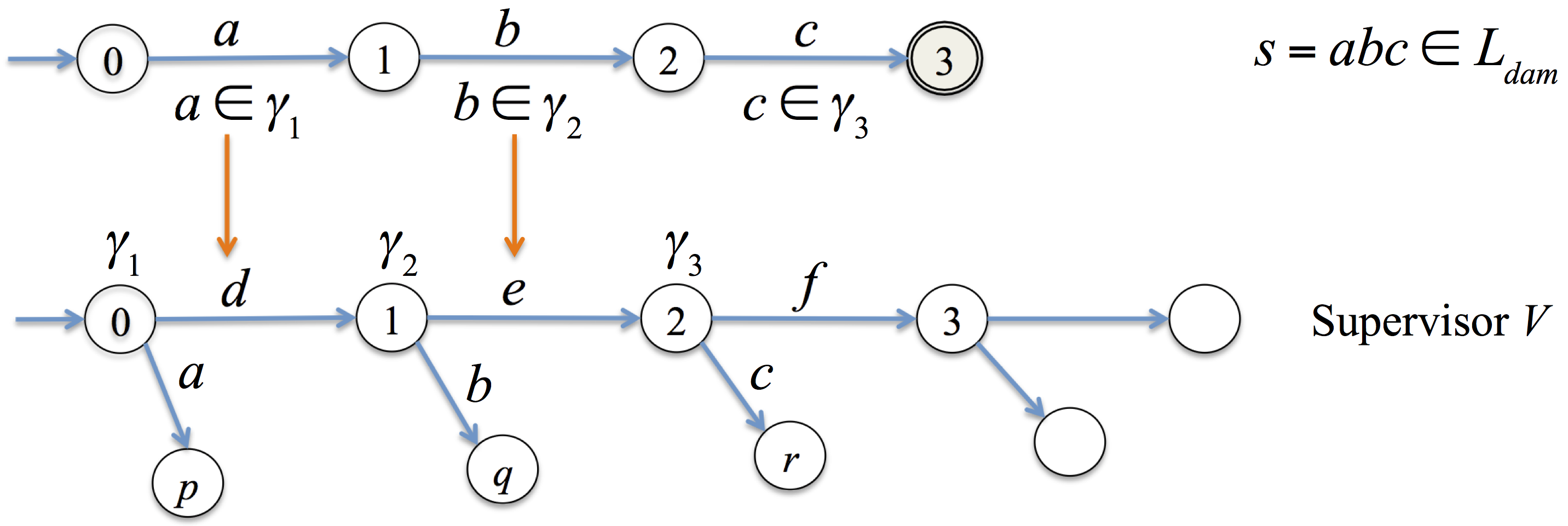}
    \end{center}
    \caption{Example 1: A smart sensor attack}
    \label{fig:Cyber-Security-100}
\end{figure}

Let us start with a small example, which is depicted in Figure \ref{fig:Cyber-Security-100}.
Assume that the attacker $A$ wants to achieve a string $abc\in L_{dam}$, which leads to a damage state. Assume that event $a$ is contained in control pattern $\gamma_1$, event $b$ is in control pattern $\gamma_2$, and event $c$ in control pattern $\gamma_3$. After event $a$ fires, the attacker wants the control pattern $\gamma_2$ to be issued. Since event $a$ does not lead to control pattern $\gamma_2$, but event $d$ does, the attacker $A$ will replace event $a$ with $d$ to trick the supervisor $S$ to generate $\gamma_2$. Assume that $b$ is fired afterwards. The attacker wants $\gamma_3$ to be issued. Since event $b$ does not lead to $\gamma_3$, instead, event $e$ does, the attacker $A$ replaces event $b$ with event $e$ to trick the supervisor $S$ to issue $\gamma_3$, if event $c$ happens afterwards, the attacker achieves his/her goal without being detected by the supervisor. The attacker could continue this trick as long as it is possible. So essentially, by faking some observable string, the attacker hopes to trick the supervisor to issue a sequence of control patterns, which contain some damaging strings, without being detected by the supervisor. We now generalize this idea. For notational simplicity, given a string $t=\nu_1\cdots\nu_n\in\Sigma^*$ with $n\in\mathbb{N}$, for each $i\in \{1,\cdots, n\}$, we use $t^i$ to denote the prefix substring $\nu_1\cdots\nu_i$. By convention, $t^0:=\epsilon$.

\begin{Theo}\label{thm0}
\textnormal{Given a plant $G$, a supervisor $V$ and a damage language $L_{dam}\subseteq L(G)-L(V/G)$, there is a smart weak sensor attack $A$, if and only if the following condition holds: there {\color{black}exists} ${\color{black}s}=u_1\sigma_1\cdots u_r\sigma_r u_{r+1}\in L_{dam}$, with $r\in\mathbb{N}$, $u_1,\cdots, u_{r+1}\in\Sigma_{uo}^*$ and $\sigma_1,\cdots,\sigma_r\in\Sigma_o$, and $t=\nu_1\cdots\nu_r\in P_o(L(V/G))$ with $\nu_1,\cdots,\nu_r\in\Delta_n$ such that (1) $u_1,\sigma_1\in V(t^0)^*$; (2) for each $i\in \{2,\cdots,r\}$, $u_i,\sigma_i\in V(t^{i-1})^*$; (3) $u_{r+1}\in V(t)^*$.\hfill $\Box$
}
\end{Theo}

As an illustration, in Example 1 depicted in Figure \ref{fig:Cyber-Security-100}, we can see that $r=3$, $\sigma_1=a$, $\sigma_2=b$, $\sigma_3=c$, $\nu_1=d$, $\nu_2=e$, $u_1=u_2=u_3=u_4=\epsilon$. 

The strings $s$ and $t$ in Theorem \ref{thm0} form a risky pair $(s,t)\in L_{dam}\times \Delta_n^*$ such that, by mapping $P_o(s)$ to $t$, the attacker can rely on the existing supervisor $V$ to inflict a weak attack on the plant $G$, without being detected by the supervisor. Since the existence of a risky pair is sufficient and necessary for the existence of a smart weak sensor attack, we will use this fact to determine the existence of a resilient supervisor. But before that,  we would like to state the following result about the decidability of the existence of a regular smart weak sensor attack.

\begin{Theo}\label{Thm2} \textnormal{Given a plant $G$, a regular supervisor $V$, and a regular damage language $L_{dam}\subseteq L(G)-L(V/G)$, the existence of a regular smart weak sensor attack is decidable.\hfill $\Box$}\end{Theo}

{\color{black}By the proof of Theorem 2 shown in the Appendix, the computational complexity of deciding the existence of a regular smart weak sensor attack is $O(|\Sigma|^2|\Delta_n|2^{|G||S||D|})$, where $S$ is an automaton realization of $V$ and $D$ is an automaton recognizing $L_{dam}$.}  

In \cite{Goes2020} the authors have shown that a deterministic attack function that ensures the covertness and weak damage infliction can always be synthesized, when it exists. But since the attack model adopted in this paper is {\color{black}different from} the one used in \cite{Goes2020}, e.g., the latter does not requires $A(\epsilon)=\epsilon$ (thus, non-existence of an attack model in our definition does not necessarily means the non-existence of an attack model in \cite{Goes2020}), and {\color{black}encodes attack moves differently}, Theorem \ref{Thm2} has its own value by providing another way of synthesizing a regular deterministic smart weak sensor attack, whenever it exists. 
%However, we do not claim it as one major contribution.   

\section{Supervisor resilient to smart sensor attacks}
In this section we explore whether there exists a sufficient and necessary condition to ensure the existence of a supervisor that is resilient to all regular smart sensor attacks, i.e., the closed-loop system is not attackable by any regular smart sensor attack. In Section 3 we have shown that there is a sufficient and necessary condition for the existence of a smart weak sensor attack shown in Theorem \ref{thm0}. Since each strong attack is also a weak attack, if we can effectively eliminate those risky pairs described in Theorem \ref{thm0}, we shall be able to prevent the existence of any smart  sensor attack. Since, given a plant $G$ and a requirement $Spec$, we can always synthesize a controllable and observable  sublanguage of $L_m(G)\cap L_m(Spec)$, without loss of generality, we assume that the plant $G$ satisfies all given requirements. Thus, we will only focus on the following problem.

\begin{Prob}Given a plant $G$ and a damage language $L_{dam}\subseteq L(G)$, synthesize a supervisor $V$ such that $V/G$ is not attackable by any regular smart sensor attack with respect to $L_{dam}$.\hfill $\Box$ \end{Prob}     

To solve this problem, we first intend to find a proper way of encoding all risky pairs. Given a string $s\in\Sigma^*$, we use $s^{\uparrow}$ to denote the last event of $s$. If $s=\epsilon$, by convention, $s^{\uparrow}:=\epsilon$. In addition, we use $s_o$ to denote the longest prefix substring of $s$, whose last event is observable, i.e., $s_o\in \overline{\{s\}}\cap (\Sigma_{uo}^*\Sigma_o)^*\cap P_o^{-1}(P_o(\{s\}))$. Thus, if $s\in\Sigma_{uo}^*$, then we can derive that $s_o=\epsilon$. 

Let $\iota:\Sigma^*\rightarrow 2^{(\Sigma\times \Gamma)^*}$ be a partial mapping, where
\begin{itemize}
\item $\iota(\epsilon):={\color{black}\{\epsilon\}}$;
\item $(\forall s\in \Sigma^*)(\forall \sigma\in\Sigma)\,\iota(s\sigma):=\iota(s)\{(\sigma,\gamma)|\sigma\in\gamma\}$.
\end{itemize}
In Example 1,  we have $(a,\gamma_1)(b,\gamma_2)(c,\gamma_3)\in\iota(abc)$. 
What the map $\iota$ does is to map each string $s\in \Sigma^*$ to a set of sequences of control patterns such that each derived control pattern sequence, say $\gamma_1\cdots\gamma_r\in\Gamma^*$, contains the string $s$ in the sense that $s\in \gamma_1\cdots\gamma_r\subseteq\Sigma^*$. By applying the map $\iota$ to the damage language $L_{dam}$, the result $\iota(L_{dam}):=\cup_{s\in L_{dam}}\iota(s)$ presents all possible sequences of control patterns, each of which contains at least one string in $L_{dam}$ - in other words, each string in $\iota(L_{dam})$ may potentially result in damage. 

\begin{figure}[htb]
    \begin{center}
      \includegraphics[width=0.45\textwidth]{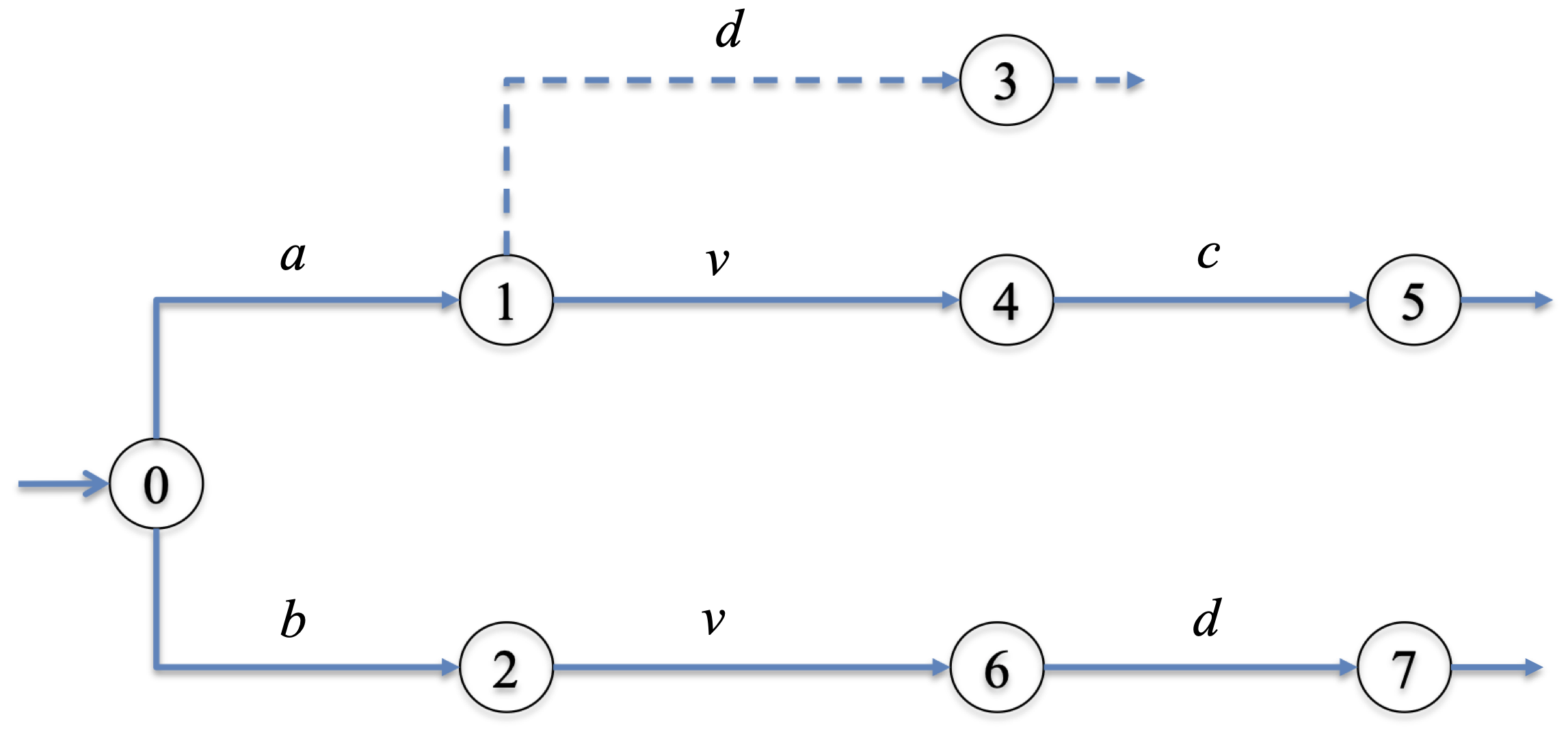}
    \end{center}
    \caption{Example 2: A small plant $G$}
    \label{fig:Cyber-Security-101}
\end{figure}

\begin{figure}[htb]
    \begin{center}
      \includegraphics[width=0.45\textwidth]{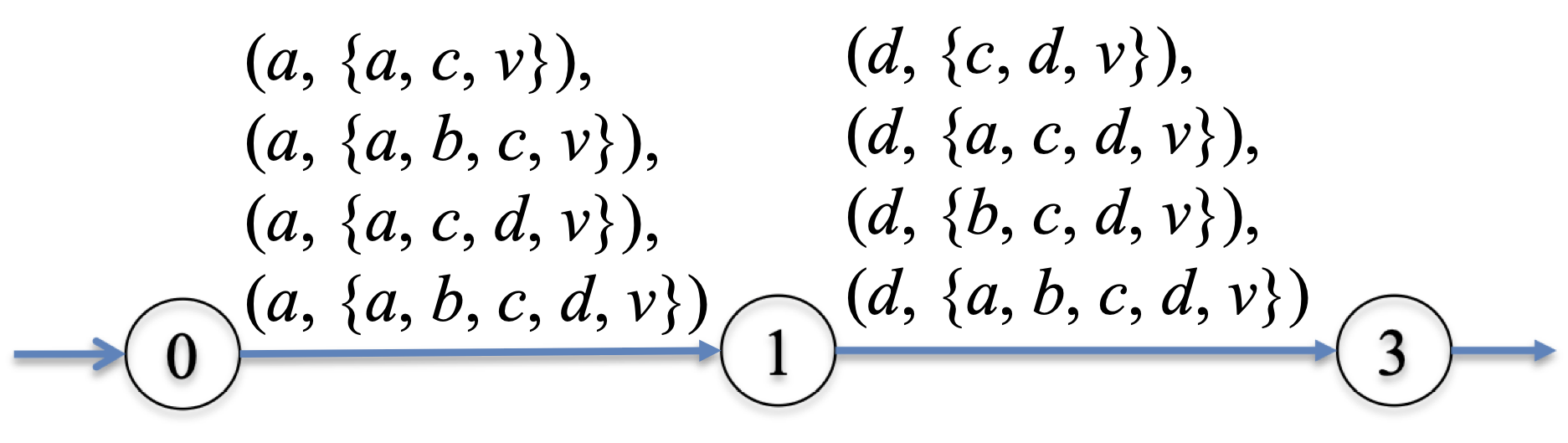}
    \end{center}
    \caption{Example 2: The model of $\iota({\color{black}L_{dam}})$}
    \label{fig:Cyber-Security-102}
\end{figure}

To further illustrate how this function works, we introduce another simple example depicted in Figure \ref{fig:Cyber-Security-101},
where $\Sigma=\{a,b,c,d,v\}$, $\Sigma_c=\{a,b,d\}$ and $\Sigma_o=\{a,b,c,d\}$. To simplify our {\color{black}illustration, in this example} we assume that $\Delta_n=\Sigma_o^{\epsilon}$, i.e., $n=1$. The damage language $L_{dam}=\{ad\}$, which is shown by a dashed line leading to state 3. Figure \ref{fig:Cyber-Security-102} depicts the outcome of $\iota({\color{black}L_{dam}})$.

A smart sensor attack replaces each intercepted observable event $\sigma\in\Sigma_o$ with a string in $\Delta_n$, unless $\sigma$ is silent, i.e., $\sigma=\epsilon$. To capture the impact of such changes on the control pattern sequences, we introduce a mapping $\psi:(\Sigma\times \Gamma)^*\rightarrow 2^{((\Sigma\cup\Delta_n)\times \Gamma)^*}$, where
\begin{itemize}
\item $\psi(\epsilon):={\color{black}\{\epsilon\}}$;
\item For each  $\mu\in (\Sigma\times\Gamma)^*$ and $(\sigma,\gamma)\in \Sigma\times \Gamma$, we have
\[\psi(\mu(\sigma,\gamma)):=\left\{\begin{array}{ll} 
\psi(\mu)\{(\sigma,\gamma)\} & \textrm{if $\sigma\in\Sigma_{uo}$,}\\
\psi(\mu)(\Delta_n\times\{\gamma\}) & \textrm{otherwise.}\end{array}\right.
\]
\end{itemize}   
We extend the domain of $\psi$ to languages in the usual way, i.e., for all $L\subseteq (\Sigma\times \Gamma)^*$, $\psi(L):=\cup_{s\in L}\psi(s)$. 

To explicitly describe how a smart attack may utilize possible sequences of control patterns, we introduce one more mapping \[\nu: ((\Sigma\cup\Delta_n)\times\Gamma)^*\rightarrow 2^{(\Sigma_o^{\epsilon}\times\Gamma)^*},\] where
\begin{itemize}
\item $\nu(\epsilon):={\color{black}\{\epsilon\}}$;
\item For all $(\sigma,\gamma)\in (\Sigma\cup\{\epsilon\})\times\Gamma$, \[\nu(\sigma,\gamma)=\left\{\begin{array}{ll} (\sigma,\gamma) & \textrm{if $\sigma\in\Sigma_o\wedge \sigma\in\gamma$;}\\
(\epsilon,\gamma) & \textrm{if $\sigma\in\Sigma_{uo}\wedge\sigma\in\gamma$;}\\
\varnothing &\textrm{otherwise.}
\end{array}\right.\]
\item For all $s=\sigma_1\cdots\sigma_r\in\Delta_n$, ${\color{black}|P_o(s)|}=r\geq 2$, and $\gamma\in\Gamma$,
\begin{center}$\nu(s,\gamma):=\{(\sigma_1,\gamma_1)\cdots (\sigma_r,\gamma)|\sigma_r\in\gamma\wedge$\\ $(\forall i\in\{1,\cdots,r-1\})\sigma_i\in \gamma_{r-1}\in\Gamma\}.$\end{center}
\item $(\forall \mu (s,\gamma)\in ((\Sigma\cup\Delta_n)\times\Gamma)^+)\, \nu(\mu (s,\gamma))=\nu(\mu)\nu(s,\gamma)$.
\end{itemize}

\begin{figure}[htb]
    \begin{center}
      \includegraphics[width=0.45\textwidth]{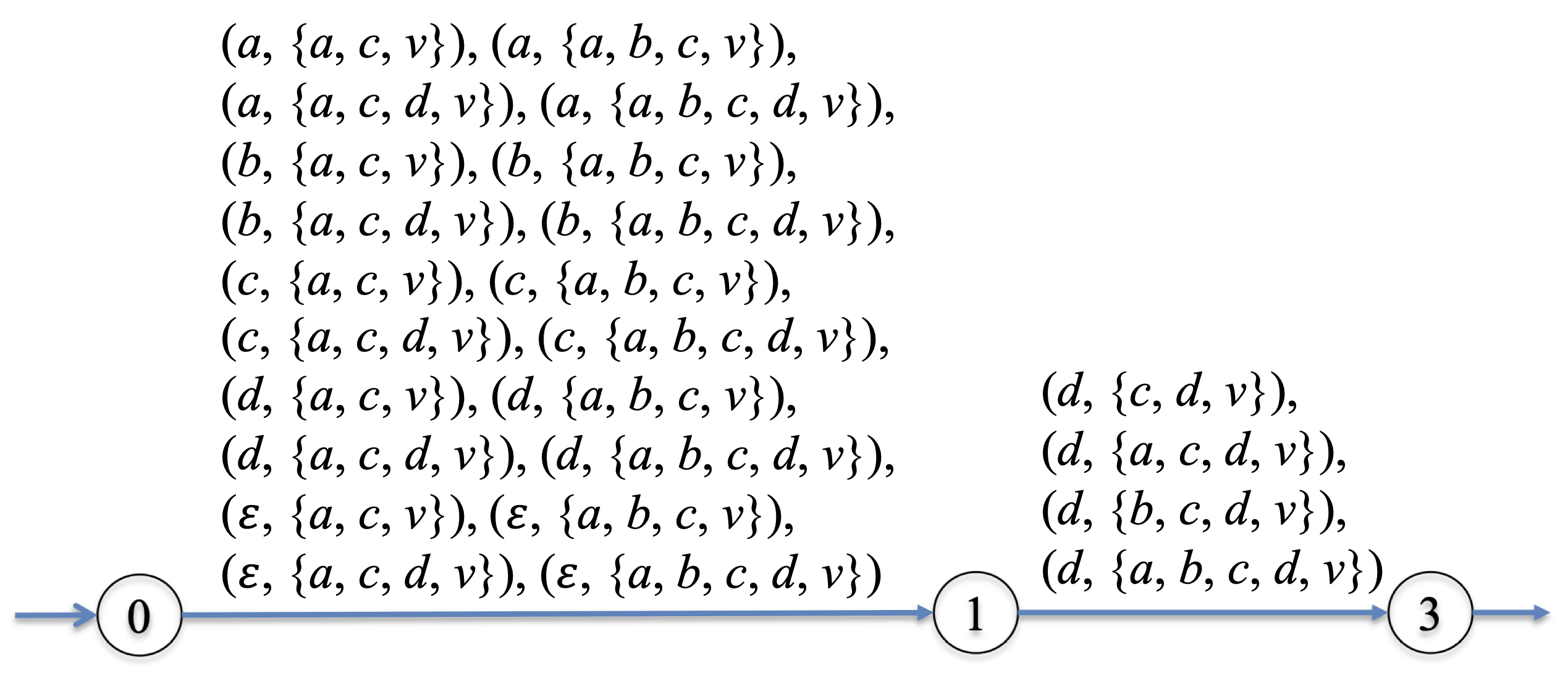}
    \end{center}
    \caption{Example 2: The model of $\psi(\iota(L(G)))$}
    \label{fig:Cyber-Security-103}
\end{figure}

As an illustration, we apply the map $\psi$ to the damage language $\iota(L_{dam})$ in Figure \ref{fig:Cyber-Security-102}, where $n=1$. {\color{black}To simplify illustration, we assume that an attacker can, but prefers not to, change events $c$ and $d$}. The outcome is depicted in Figure \ref{fig:Cyber-Security-103}. Notice that when event $a$ is intercepted by the attacker, it can be replaced by any other strings in $\Delta_n$. Because $n=1$, we have $\Delta_1=\Sigma_o\cup\{\epsilon\}=\{a,b,c,d,\epsilon\}$ - in this case, the outcome of $\nu(\psi(\iota(L(G))))$ equals $\psi(\iota(L(G)))$. 

Next, we determine all control pattern sequences in the plant $G$ that may be used by a smart attack. Let $\Upsilon_G:L(G)\times\Sigma^*\times\Gamma\rightarrow \{0,1\}$ be a Boolean map, where for each $(s,t,\gamma)\in L(G)\times\Sigma^*\times\Gamma$, 
\[\Upsilon_G(s,t,\gamma)=1\iff st\in  L(G)\wedge t\in \gamma^*.\]
For each $\gamma\in\Gamma$, let \[B(\gamma):=\left\{\begin{array}{ll} \{(\epsilon,\gamma)\}^* & \textrm{if $\gamma\cap\Sigma_{uo}\neq\varnothing$,}\\
\{\epsilon\} & \textrm{otherwise.}\end{array}\right.\]

%\[\bigcup_{\gamma\in\Gamma:\gamma\cap\Sigma_{uo}\neq\varnothing}\{(\epsilon,\gamma)\}^*(\epsilon \cup \{(\epsilon,\gamma)\in \{\epsilon\}\times\Gamma| \gamma\subseteq\Sigma_o\}).\] 

\noindent Let $p:(\Sigma_o^{\epsilon}\times\Gamma)^*\rightarrow\Gamma^*$ be a projection map, where
\begin{itemize}
\item $p(\epsilon):=\epsilon$;
\item $(\forall s(\sigma,\gamma)\in (\Sigma_o^{\epsilon}\times\Gamma)^+)\, p(s(\sigma,\gamma)):=p(s)\gamma$.
\end{itemize}
\noindent And let $g:(\Sigma_o^{\epsilon}\times\Gamma)^*\rightarrow \Sigma_o^*$ be a projection map, where
\begin{itemize}
\item $g(\epsilon):=\epsilon$;
\item $(\forall \mu (\sigma,\gamma)\in (\Sigma_o^{\epsilon}\times\Gamma)^+)\, g(\mu(\sigma,\gamma)):=g(\mu)P_o(\sigma)$.
\end{itemize}

\noindent Let $\zeta:L(G)\rightarrow 2^{(\Sigma_o^{\epsilon}\times \Gamma)^*}$ be a total mapping, where
\begin{itemize}
\item $\zeta(\epsilon):=(\cup_{\gamma\in\Gamma:\gamma\cap\Sigma_{uo}\neq\varnothing}\{(\epsilon,\gamma)\}^+)\bigcup (\cup_{\gamma\in\Gamma:\gamma\subseteq\Sigma_o}\{(\epsilon,\gamma)\})$;
\item For all $s\in (\Sigma_{uo}^*\Sigma_o)^*$ and $t\in \Sigma_{uo}^*\Sigma_o^{\epsilon}$ with $st\in L(G)$, \[\zeta(st):={\color{black}\left\{\begin{array}{ll}\zeta(s) & \textrm{if $P_o(t)=\epsilon$,}\\ M & \textrm{otherwise,}\end{array}\right.}\]
where \[M:=
\bigcup_{w\in \zeta(s):\Upsilon_G(s,t,p(w)^{\uparrow})=1}\bigcup_{\gamma'\in\Gamma}\{w(P_o(t),\gamma')\}B(\gamma').
\]
\end{itemize}
We call $\zeta(L(G))$ the {\em augmented closed behaviour} of $G$. The {\em augmented marked behaviour} of $G$ induced by $\zeta$ is defined as $\zeta(L(G))\cap g^{-1}(P_o(L_m(G)))$. 

This definition of $\zeta$ indicates that, except for control patterns generated initially, i.e., when $s=\epsilon$, each  control pattern will be changed only after an observable event is received, i.e., when $st\in \Sigma^*\Sigma_o\cap L(G)$. This matches the definition of a supervisor $V$ that changes its output only when a new observation is received. In addition, if a control pattern $\gamma$ contains unobservable events, it will be contained in a self-loop of the augmented event $(\epsilon,\gamma)$, i.e., $\{(\epsilon,\gamma)\}^*$, denoting that the control pattern $\gamma$ may be used more than once by the plant, as long as no new observable event has been received. Again, this matches the Ramadge-Wonham supervisory control paradigm, where execution of any unobservable transition allowed by the current control pattern will not change the current control pattern - recall that in a finite-state automaton realization of $V$, unobservable events are self-looped at relevant states.      

\begin{figure}[htb]
    \begin{center}
      \includegraphics[width=0.45\textwidth]{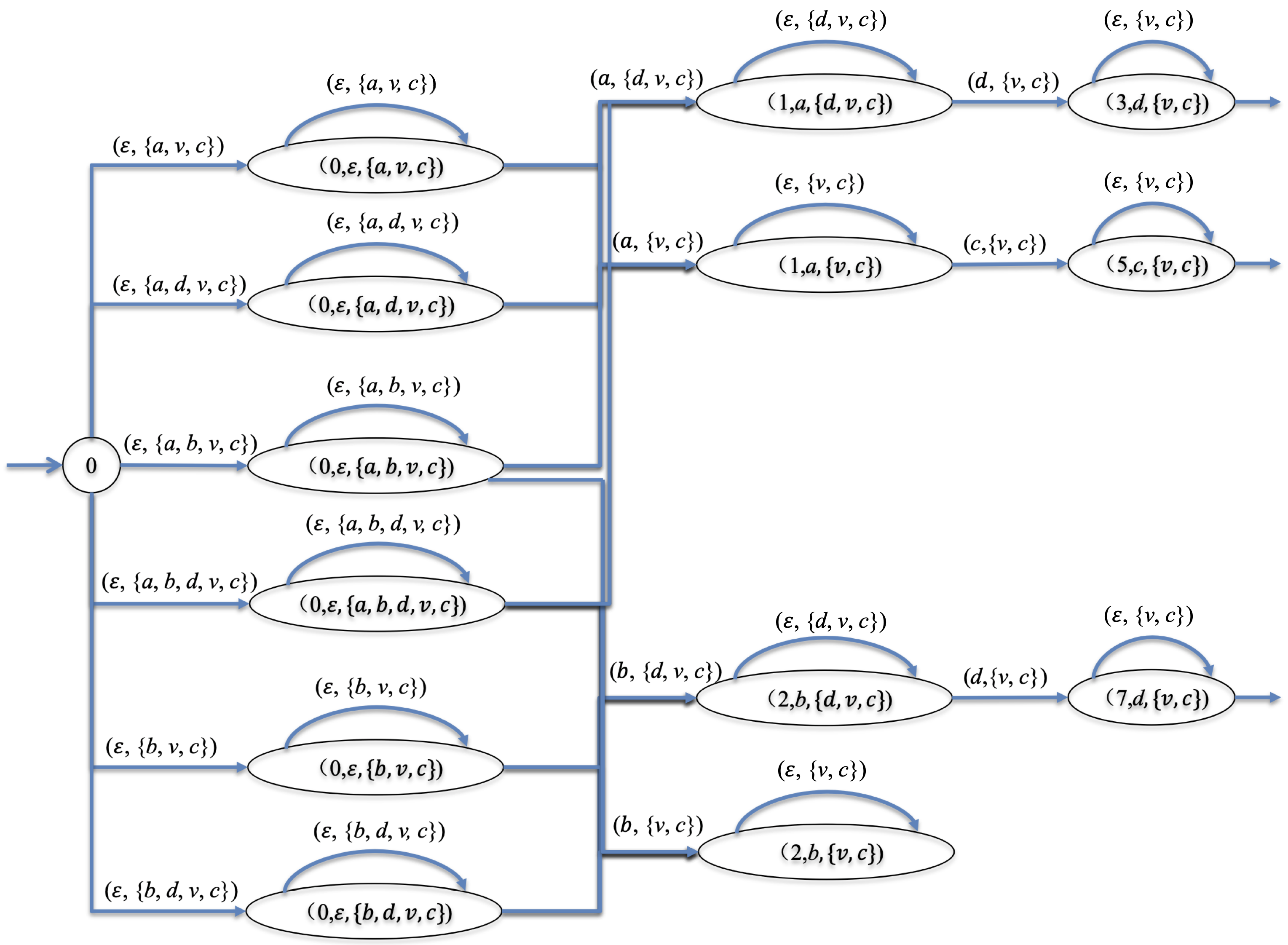}
    \end{center}
    \caption{Example 2: The model of $\zeta(L(G))$}
    \label{fig:Cyber-Security-104}
\end{figure}

As an illustration, we apply $\zeta$ to the plant model $L(G)$ depicted in Figure \ref{fig:Cyber-Security-101}. Part of the outcome is depicted in Figure \ref{fig:Cyber-Security-104}.  
Because the total state set is $X\times\Sigma_o^{\epsilon}\times\Gamma$, which is too big to be shown entirely in the picture, we only include states that have at least one future extension, unless they are marker states {\color{black}(i.e., states $(3,d,\{v,c\})$, $(5,c,\{v,c\})$ and $(7,d,\{v,c\})$)}, except for one blocking state $(2,b,\{v,c\})$, which is left there for an illustration purpose that will be explained shortly. The marker states in   Figure \ref{fig:Cyber-Security-104} denote the augmented marked behaviour of $G$ in Example 2.

Till now, we have provide sufficient means to describe all risky pairs, which are captured by $\nu(\psi(\iota(L_{dam})))$ at the attacker's demand side, and $\zeta(L(G))$ at the plant's supply side. To avoid such risky pairs, we only need to remove $p^{-1}(p(\nu(\psi(\iota(L_{dam}))))(\Sigma_o^{\epsilon}\times\Gamma)^*$ from $\zeta(L(G))$. The reason why we concatenate $(\Sigma_o^{\epsilon}\times\Gamma)^*$ at the end of $p^{-1}(p(\nu(\psi(\iota(L_{dam}))))$ is to denote all possible augmented strings that may contain some strings in $p^{-1}(p(\nu(\psi(\iota(L_{dam}))))$ as prefix substrings. Thus, all safe supervisory control pattern sequences shall be contained in $\hat{H}:=\zeta(L(G))-p^{-1}(p(\nu(\psi(\iota(L_{dam}))))(\Sigma_o^{\epsilon}\times\Gamma)^*$ in order to prevent any sequence of control patterns from being used by an attacker.  

\begin{figure}[htb]
    \begin{center}
      \includegraphics[width=0.45\textwidth]{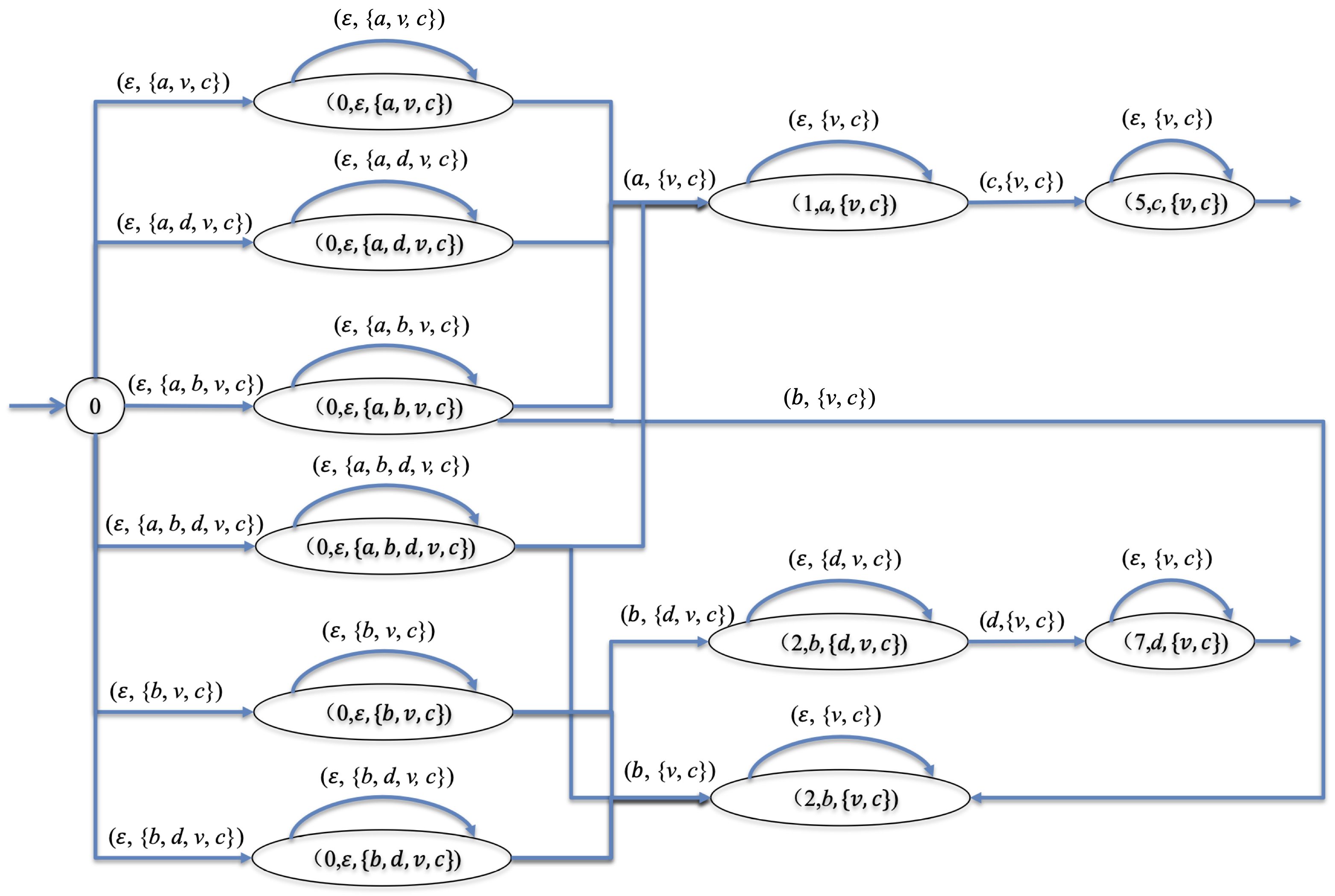}
    \end{center}
    \caption{Example 2: The model of $\hat{H}$}
    \label{fig:Cyber-Security-105}
\end{figure}

Figure \ref{fig:Cyber-Security-105} depicts   
the outcome of subtracting risky control pattern sequences $p^{-1}(p(\nu(\psi(\iota(L_{dam}))))(\Sigma_o^{\epsilon}\times\Gamma)^*$ from (part of) $\zeta(L(G))$ shown in Figure \ref{fig:Cyber-Security-104}. It is clear that there cannot be any sequence of control patterns $\gamma_1\gamma_2\cdots\gamma_l$ such that $a\in\gamma_1$ and $d\in\gamma_2$.

To extract a proper supervisor from $\hat{H}$, we need a few more  technical preparations. Let $H:=\overline{\hat{H}}\subseteq (\Sigma_o^{\epsilon}\times\Gamma)^*$ be the prefix closure of $\hat{H}$. 
%\begin{Def}\label{Def1}
%\textnormal{A sublanguage $L\subseteq H$ is {\em observable} with respect to $\zeta(L(G))$ and $g$, if for all $s,s'\in \overline{L}$ and $(\sigma,\gamma)\in \Sigma_o^{\epsilon}\times\Gamma$, we have that $s(\sigma,\gamma)\in\overline{L}$, $s'(\sigma,\gamma)\in \zeta(L(G))$ and $g(s)=g(s')$ imply that $s'(\sigma,\gamma)\in\overline{L}$.\hfill $\Box$}\end{Def}
All tuples $(\sigma,\gamma)\in\Sigma_o\times\Gamma$ are considered to be controllable, except for tuples $\{\epsilon\}\times\Gamma$. We introduce the concept of conditional controllability inspired by the standard notion of controllability in \cite{WR87}.

\begin{Def}\label{Def2}
\textnormal{A sublanguage $L\subseteq H$ is {\em conditionally controllable} with respect to $\zeta(L(G))$ and $\{\epsilon\}\times\Gamma$, if 
\begin{flushright}$(\overline{L}-\{\epsilon\})(\{\epsilon\}\times\Gamma)\cap \zeta(L(G))\subseteq\overline{L}.\hspace*{1cm}\Box$\end{flushright}
}
\end{Def}
In other words, as long as $(\epsilon,\gamma)$ is not defined at the beginning, i.e., $(\epsilon,\gamma)\notin\zeta(L(G))$, it should not be disabled, if it follows a non-empty string $s\in\overline{L}$. We can briefly explain the  motivation as follows. If an event $(\epsilon,\gamma)$ does not appear at the beginning, by the definition of $\zeta(L(G))$ and subsequently that of $H$, it must be incurred by another string $s(\sigma,\gamma)$ such that $\gamma\cap\Sigma_{uo}\neq\varnothing$ -- clearly, we can stop $(\sigma,\gamma)$, if $\sigma\neq\epsilon$, by not choosing $\gamma$; but after $\gamma$ is chosen and some unobservable event allowed by $\gamma$ occurs, the same control pattern $\gamma$ will continuously remain active, i.e., $(\epsilon,\gamma)$ will still be allowed, until a new observation is generated, leading to a new control pattern. But the situation is different initially, as we can directly disable the control pattern $\gamma$, thus stop the event $(\epsilon,\gamma)$. It is clear that conditional controllability is also closed under set union. 

Let $\mathcal{C}(\zeta(L(G)),H)$ be the set of all prefix-closed sublanguages of $H$, which is conditionally controllable with respect to $\zeta(L(G))$ and $\{\epsilon\}\times\Gamma$. It is clear that the supremal conditionally controllable sublanguage in $\mathcal{C}(\zeta(L(G)),H)$ under the partial order of set inclusion exists. We denote this unique sublanguage as $\mathcal{S}_*:=\textrm{sup}\mathcal{C}(\zeta(L(G)),H)$. Notice that $\mathcal{S}_*$ contains no sequence of control patterns that may be used by a smart attack to inflict damage. Later, we will show that $\mathcal{S}_*$ contains all feasible supervisors that are resilient to smart sensor attacks, as long as such a supervisor exists. We now introduce techniques to extract a feasible resilient supervisor out of $\mathcal{S}_*$, if it exists. To this end, we introduce a few more concepts.

Let $f:\mathcal{S}_*\rightarrow 2^X$ be a mapping, where
\begin{itemize}
\item For all $(\epsilon,\gamma)\in\mathcal{S}_*$,\[f(\epsilon,\gamma):=\{x\in X|(\exists t\in\gamma^*\cap \overline{\Sigma_{uo}^*\Sigma_o})\xi(x_0,t)=x\};\]
\item For all $s\in \mathcal{S}_*$ and $(\sigma,\gamma)\in \Sigma_o^{\epsilon}\times\Gamma$ with $s(\sigma,\gamma)\in \mathcal{S}_*$, if $\sigma=\epsilon$, then $f(s(\sigma,\gamma)):=f(s)$; otherwise,
\begin{center} $f(s(\sigma,\gamma)):=$\\ $\{x\in X|(\exists t\in\gamma^*\cap \overline{\Sigma_{uo}^*\Sigma_o})(\exists x'\in f(s))\xi(x',\sigma t)=x\}.$\end{center}
\end{itemize}   

The map $f$ essentially associates each string $s\in\mathcal{S}_*$ with the corresponding state estimate of $G$. Let $h:\mathcal{S}_*\rightarrow 2^X$ be the marking coreachability map associated with the plant $G$, where for each $s=(\epsilon,\gamma_0)(\sigma_1,\gamma_1)\cdots (\sigma_n,\gamma_n)\in\mathcal{S}_*$ with $n\in\mathbb{N}$,
\begin{itemize}
\item $f(s)\cap X_m=\varnothing\Rightarrow h(s)=\varnothing$;
\item If $f(s)\cap X_m\neq\varnothing$, then let  $\varrho:2^X\times\Gamma\rightarrow 2^X$,
where for all $U\in 2^X$ and $\gamma\in\Gamma$, 
\begin{center}$\varrho(U,\gamma):=$\\ $\{x\in X|(\exists t\in\gamma^*\cap \overline{\Sigma_{uo}^*\Sigma_o})(\exists x'\in U)\xi(x,t)=x'\},$\end{center}
and $h(s):= \cup_{i=0}^n U_i$, where
\begin{itemize}
\item $U_n:=\varrho(f(s)\cap X_m,\gamma_n)$;
\item $(\forall i\in \{0,\cdots,n-1\})\, U_i:=\varrho(U_{i+1},\gamma_i)$.
\end{itemize}
\end{itemize} 

\begin{Def}\label{Def5}
\textnormal{A resilient supervisor candidate $\mathcal{L}\subseteq \mathcal{S}_*$ is {\em nonblocking} with respect to $G$, if for all $s\in \mathcal{L}$,
\[f(s)\subseteq \bigcup_{t\in \mathcal{L}:s\leq t}h(t).\] 
\hfill $\Box$}
\end{Def}

\begin{Def}\label{Def4}\textnormal{A sublanguage $\mathcal{L}\in \mathcal{C}(\zeta(L(G)),H)$ is a {\em nonblocking resilient supervisor candidate} 
if for all $s\in \mathcal{L}$,
\begin{enumerate}
\item 
$(\forall t\in g^{-1}(g(s))\cap \mathcal{L}) p(t)^{\uparrow}=p(s)^{\uparrow}$;
\item {\color{black}$g(En_{\mathcal{L}}(s))= P_o(p(s^{\uparrow}))\cap g(En_{\zeta(L(G))}(s))$};
\item $\mathcal{L}$ is nonblocking with respect to $G$.\hfill $\Box$
\end{enumerate}
}\end{Def} 
Notice that $t\in g^{-1}(g(s))\cap \mathcal{L}$ means that $g(s)=g(t)$, and $ p(t)^{\uparrow}=p(s)^{\uparrow}$ means that the incurred control patterns by $g(t)$ and $g(s)$ are the same. Thus,  the first condition in Def. \ref{Def4} essentially states that, all observably identical strings must lead to the same control pattern -- consequently, any silent transition $\epsilon$ cannot generate any new control pattern other than the current one. The second condition states that {\color{black}an ``observable'' event $\sigma\in\Sigma_o^\epsilon$ is allowed by $\mathcal{L}$, i.e., $\sigma\in g(En_{\mathcal{L}}(s))$, if and only if it is allowed by the  control pattern incurred by $s$, i.e., $\sigma\in P_o(p(s^{\uparrow}))$, and also allowed by $L(G)$, i.e., $\sigma\in g(En_{\zeta(L(G))}(s)))$.} The last condition refers to nonblockingness of $\mathcal{L}$.

\begin{figure}[htb]
    \begin{center}
      \includegraphics[width=0.45\textwidth]{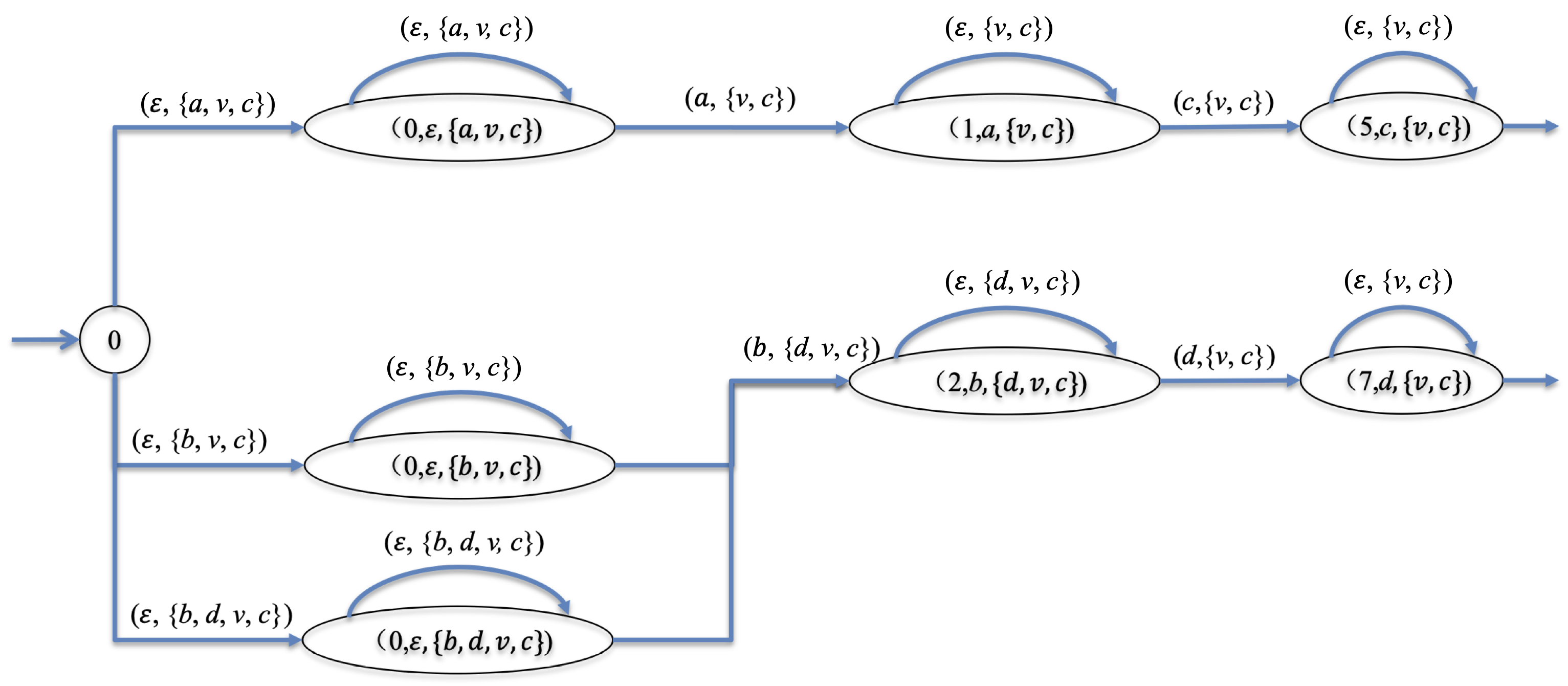}
    \end{center}
    \caption{Example 2: The set of all nonblocking resilient supervisor candidates of $\mathcal{S}_*$}
    \label{fig:Cyber-Security-106}
\end{figure}

As an illustration, we calculate $\textrm{sup}\mathcal{C}(\zeta(L(G)),H)$ and remove all states that {\color{black}violate either one of the conditions} of Def. \ref{Def4}. Figure \ref{fig:Cyber-Security-106} depicts   
the outcome. We can see that the state $(2,b,\{v,c\})$ in Figure \ref{fig:Cyber-Security-105} needs to be removed because it is blocking, violating the third condition in Def. \ref{Def4}. In addition, states $(0,\epsilon,\{a,b,v,c\})$ and $(0,\epsilon,\{a,b,d,v,c\})$ and $(0,\epsilon,\{a,d,v,c\})$ in Figure \ref{fig:Cyber-Security-105} also need to be removed because they clearly violate the second condition of Def. \ref{Def4}, as the event $b$ is defined in control patterns $\{a,b,v,c\}$ and $\{a,b,d,v,c\}$ of states $(0,\epsilon,\{a,b,v,c\})$ and $(0,\epsilon,\{a,b,d,v,c\})$, respectively, but no outgoing transitions containing $b$ are allowed at these two states in $H$, even though these transitions are allowed in $\zeta(L(G))$, and event $d$ is defined in the control pattern $\{a,d,v,c\}$ of state $(0,\epsilon,\{a,d,v,c\})$, but no outgoing transition containing $d$ is allowed in $H$, even though such a transition is allowed in $\zeta(L(G))$. 

We now state the following theorem, which is the first step towards solving the decidability problem of the existence of a supervisor resilient to smart sensor attacks.

\begin{Theo}\label{Thm3}
\textnormal{Given a plant $G$ and a damage language $L_{dam}\subseteq L(G)$, let $\mathcal{S}_*$ be defined above. Then there exists a supervisor $V:P_o(L(G))\rightarrow\Gamma$ such that $V/G$ is not attackable w.r.t. $L_{dam}$, iff there exists a nonblocking resilient supervisor candidate $\mathcal{L}\subseteq \mathcal{S}_*$.\hfill $\Box$}
\end{Theo}

As an illustration, we can check that any marked sequence in Figure \ref{fig:Cyber-Security-106} is a nonblocking resilient supervisor candidate. For example, take a look at the sublanguage $\mathcal{L}:=\{(\epsilon,\{a,v,c\})\}^+\{(a,\{v,c\})\}\{(\epsilon,\{v,c\})\}^*$ $\{(c,\{v,c\})\}\{(\epsilon,\{v,c\})\}^*$. We can check that $\mathcal{L}$ is conditional controllable with respect to $\zeta(L(G))$ and $\{\epsilon\}\times\Gamma$. Thus, $\mathcal{L}\in\mathcal{C}(\zeta(L(G)),H)$. In addition, $\mathcal{L}$ is nonblocking and satisfies conditions  in Def. \ref{Def4}. Thus, $\mathcal{L}$ is a nonblocking resilient supervisor candidate of $\mathcal{S}_*$. By Theorem \ref{Thm3}, we know that there must exist a resilient supervisor $V$ that does not allow any smart sensor attack.  Based on the construction shown in the proof of Theorem \ref{Thm3}, the corresponding supervisor is $V(\epsilon):=\{a,v,c\}$, $V(a):=\{v,c\}$ and $V(ac):=\{v,c\}$. For any other observable string $s\in P_o(L(G))$, we simply set $V(s):=\Sigma_{uc}$. 

Theorem \ref{Thm3} indicates that, to decide whether there exists a nonblocking supervisor that disallows smart sensor attacks, we only need to decide whether there exists a nonblocking resilient supervisor candidate $\mathcal{L}\subseteq\mathcal{S}_*$. Next, we shall discuss how to determine the existence of such a language $\mathcal{L}$.

\section{Decidability of the existence of a supervisor resilient to smart sensor attacks}

In the previous section we present a sufficient and necessary condition for the existence of a resilient supervisor. However, the computability issue is not addressed. In this section, we discuss how to compute all those sets and languages introduced in the previous section, and eventually show how to decide the existence of a resilient supervisor, i.e., to decide when that sufficient and necessary condition mentioned in Theorem \ref{Thm3} holds for a given plant $G$ and a regular damage language $L_{dam}$.  

We first discuss how to compute $\iota(L_{dam})$. As shown in Section 4, let $D=(W,\Sigma,\kappa,w_0,W_m)$ recognize $L_{dam}$, i.e., $L_m(D)=L_{dam}$. We construct another finite-state automaton $D_{\iota}:=(W,\Sigma\times\Gamma,\kappa_\iota,w_0, W_m)$, where $\kappa_\iota:W\times\Sigma\times\Gamma\rightarrow W$ is the (partial) transition map such that for each $(w,\sigma,\gamma)\in W\times\Sigma\times\Gamma$ and $w'\in W$,
\[\kappa_\iota(w,\sigma,\gamma)=w'\iff \sigma\in\gamma\wedge \kappa(w,\sigma)=w'.\]

\begin{Pro}\label{Prop3}
\textnormal{$\iota(L_{dam})=L_m(D_\iota)$.\hfill $\Box$}
\end{Pro}
Proof: It is clear from the construction of $D_\iota$.\hfill $\blacksquare$

Next, we describe how to calculate $\psi(\iota(L_{dam}))$. Let $D_{\psi}=(W,(\Sigma\cup \Delta_n)\times\Gamma,\kappa_{\psi},w_0, W_m)$, where $\kappa_\psi:W\times (\Sigma\cup\Delta_n)\times\Gamma\rightarrow W$ is the (partial) transition map such that for each $(w,u,\gamma)\in W\times (\Sigma\cup\Delta_n)\times\Gamma$ and $w'\in W$, we have $\kappa_\psi(w,u,\gamma)=w'$ if one of the following holds:
\begin{itemize}
\item $u\in\Sigma_{uo}\,\wedge\, \kappa_\iota(w,u,\gamma)=w'$;
\item $u\in\Delta_n\,\wedge\, (\exists \sigma\in\Sigma_o)\,\kappa_\iota(w,\sigma,\gamma)=w'$.
\end{itemize}

\begin{Pro}\label{Prop4}
\textnormal{$\psi(\iota(L_{dam}))=L_m(D_{\psi})$.\hfill $\Box$}
\end{Pro}
Proof: By the construction of $D_\psi$ and the definition of $\psi$, the proposition follows. \hfill $\blacksquare$, 

Next, we describe how to calculate $\nu(\psi(\iota(L(G))))$ by modifying $D_\psi$. For each transition $\kappa_\psi(w,u,\gamma)=w'$, if $u\in\Delta_n$ and $|u|\geq 2$, we make the following changes to $D_\psi$. Assume that $u=\sigma_1\cdots\sigma_r$ with $r\in\mathbb{N}$ and $\sigma_i\in\Sigma_o$ ($i\in\{1,\cdots,r\}$). We create $r-1$ new states $\tilde{w}_1,\cdots,\tilde{w}_{r-1}$ such that for  each sequence $\gamma_1\cdots\gamma_{r-1}\in\Gamma^*$ with $\sigma_i\in\gamma_i$ ($i=1,\cdots,r$), we define $\kappa_\psi (w,\sigma_1,\gamma_1)=\tilde{w}_1$, $\kappa_\psi(\tilde{w}_i,\sigma_{i+1},\gamma_{i+1})=\tilde{w}_{i+1}$ ($i=1,\cdots,r-2$) and $\kappa_\psi(\tilde{w}_{r-1},\sigma_r,\gamma)=w'$. Add newly created states to the state set $W$ of $D_\psi$ and new transitions to $\kappa_\psi$. Continue this process until all transitions are processed. Let the final finite-state automaton be $D_\nu$.    

\begin{Pro}\label{Prop4-1}
\textnormal{$\nu(\psi(\iota(L_{dam})))=L_m(D_{\nu})$.\hfill $\Box$}
\end{Pro}
Proof: By the construction of $D_\nu$ and the definition of $\nu$, the proposition follows. \hfill $\blacksquare$

Next, we will show how to compute $\zeta(L(G))$. We construct a nondeterministic finite-state automaton $G_\zeta:=(X\times\Sigma_o^{\epsilon}\times\Gamma,\Sigma_o^{\epsilon}\times\Gamma,\xi_\zeta,(x_0,\epsilon,\Sigma), X_m\times\Sigma_o^{\epsilon})$, where \[\xi_\zeta: X\times\Sigma_o^{\epsilon}\times\Gamma\times \Sigma_o^{\epsilon}\times\Gamma\rightarrow 2^{X\times\Sigma_o^{\epsilon}\times\Gamma}\] is the nondeterministic transition map such that
\begin{itemize}
\item For all $\gamma\in\Gamma$,
$\xi_\zeta(x_0,\epsilon,\Sigma,\epsilon,\gamma):=\{(x_0,\epsilon,\gamma)\}$;
\item For all $(x,\sigma,\gamma)\in X\times\Sigma_o^{\epsilon}\times\Gamma-\{ (x_0,\epsilon,\Sigma)\}$,  and $(\sigma',\gamma')\in\Sigma_o^{\epsilon}\times\Gamma$, we have that
\begin{itemize}
\item if $\sigma'=\epsilon$ and $\gamma'=\gamma$ and $\gamma\cap\Sigma_{uo}\neq\varnothing$, then 
\[\xi_\zeta(x,\sigma,\gamma,\epsilon,\gamma)=\{(x,\sigma,\gamma)\};\]
\item if $\sigma'\in\Sigma_o$, then 
\begin{center}$\xi_\zeta(x,\sigma,\gamma,\sigma',\gamma'):=\{(x',\sigma',\gamma')\in X\times\Sigma_o^{\epsilon}\times\Gamma |$\\ $(\exists u\in P_o^{-1}(\sigma')\cap \gamma^*\cap (\Sigma_{uo}^*\Sigma_o)^*)\xi(x,u)=x'\}.$\end{center}
\end{itemize}
\end{itemize}

\begin{figure}[htb]
    \begin{center}
      \includegraphics[width=0.45\textwidth]{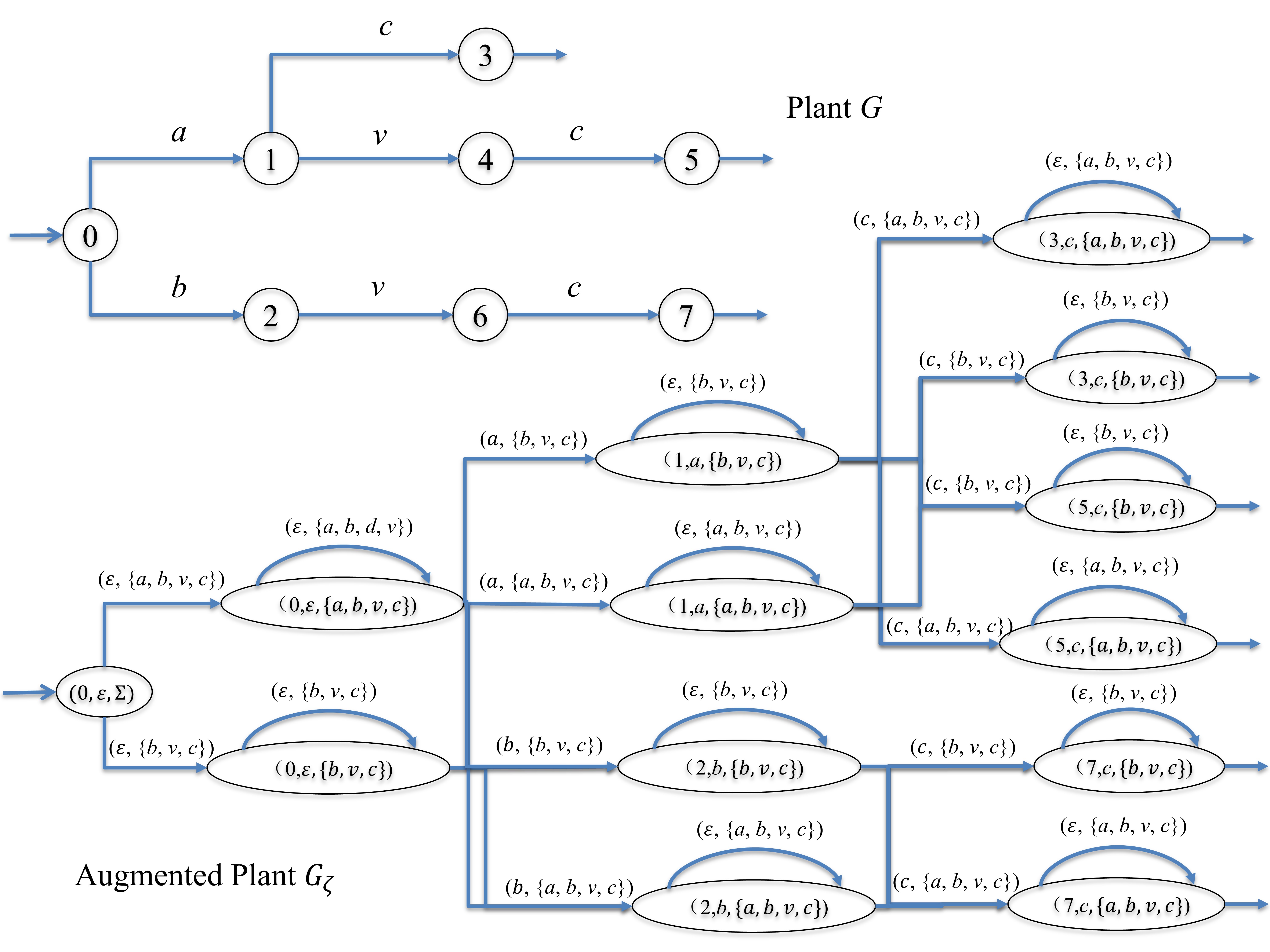}
    \end{center}
    \caption{Example 3: A plant $G$ and the corresponding $G_\zeta$}
    \label{fig:Cyber-Security-107}
\end{figure}

To illustrate the construction procedure for $G_\zeta$, a small example is depicted in Figure \ref{fig:Cyber-Security-107},
where $\Sigma=\{a,b,c,v\}$, $\Sigma_c=\{a\}$ and $\Sigma_o=\{a,b,c\}$. Thus, there are only two control patterns $\gamma_1=\{a,b,v,c\}$ and $\gamma_2=\Sigma_{uc}=\{b,v,c\}$. The outcome of $G_\zeta$ is shown in Figure \ref{fig:Cyber-Security-107}, where nondeterministic transitions occur at both (augmented) states $(1,a,\{b,v,c\})$ and $(1,a,\{a,b,v,c\})$. 

\begin{Pro}\label{Prop5}
\textnormal{$\zeta(L(G))=L(G_{\zeta})$.\hfill $\Box$}
\end{Pro}
Proof: By the definition of $\zeta$ and the construction of $G_\zeta$, it is clear that $\zeta(L(G))\subseteq L(G_\zeta)$. So we only need to show that $L(G_\zeta)\subseteq \zeta(L(G))$. We use induction. At the initial state $(x_0,\epsilon,\Sigma)$, for each $\gamma\in\Gamma$, if $\gamma\cap\Sigma_{uo}\neq\varnothing$, we have $\kappa_\zeta(x_0,\epsilon,\Sigma,\epsilon,\gamma)=\{(x_0,\epsilon,\gamma)\}$ and $\kappa_\zeta(x_0,\epsilon,\gamma,\epsilon,\gamma)=\{(x_0,\epsilon,\gamma)\})$, namely $\{(\epsilon,\gamma)\}^+\subseteq L(G_\zeta)$. By the definition of $\zeta(L(G))$, we know that $\{(\epsilon,\gamma)\}^+\subseteq \zeta(L(G))$. If $\gamma\cap\Sigma_{uo}=\varnothing$, then we have $\kappa_\zeta(x_0,\epsilon,\Sigma,\epsilon,\gamma)=\{(x_0,\epsilon,\gamma)\}$, namely $(\epsilon,\gamma)\in L(G_\zeta)$.  By the definition of $\zeta(L(G))$, we know that $(\epsilon,\gamma)\in \zeta(L(G))$. Thus, the base case holds. Assume that $s\in\zeta(L(G))\cap L(G_\zeta)$, and $s(\sigma,\gamma)\in L(G_\zeta)$, we need to show that  $s(\sigma,\gamma)\in \zeta(L(G))$. If $\sigma=\epsilon$, then since $s(\sigma,\gamma)\in L(G_\zeta)$, we know that $\gamma=p(s)^{\uparrow}$ and $\gamma\cap\Sigma_{uo}\neq\varnothing$. Since $s\in \zeta(L(G))$ and $p(s)^{\uparrow}=\gamma$ and $\gamma\cap\Sigma_{uo}\neq\varnothing$, we know that $s(\epsilon,\gamma)\in \zeta(L(G))$. If $\sigma\in\Sigma_o$, then clearly there exists $tu\in L(G)$ such that $g(s)=P_o(t)$ and $u\in P_o^{-1}(\sigma)\cap p(s)^*\cap (\Sigma_{uo}^*\Sigma_o)^*$. Clearly, $s\in\zeta(t)$ and $P_o(u)=\sigma\neq\epsilon$. Thus, by the definition of $\zeta(L(G))$, we know that $s(P_o(u),\gamma)=s(\sigma,\gamma)\in \zeta(L(G))$. Thus, the induction holds, which completes the proof. \hfill $\blacksquare$

Notice that in $G_\zeta$, except for being at the initial state $(x_0,\epsilon,\Sigma)$, no transition between two different states can be unobservable. 

Since the map $p$ introduced before is a projection, it is not difficult to check that $\hat{H}=\zeta(L(G))-p^{-1}(p(\nu(\psi(\iota(L_{dam})))))(\Sigma_o^{\epsilon}\times\Gamma)^*$ is regular, as both $\zeta(L(G))$ and $\nu(\psi(\iota(L_{dam})))$ are shown to be regular. Thus, its prefix closure $H:=\overline{\hat{H}}$ is also regular. Let the alphabet be $\Sigma_o^{\epsilon}\times\Gamma$ and the uncontrollable alphabet be $\{\epsilon\}\times\Gamma$. Since $G_\zeta$ is nondeterministic, $H$ can be recognized by a nondeterministic automaton, without masking out necessary marking information inherited from $G$, which will be used later. By using a synthesis algorithm similar to the one proposed in \cite{SSR10} \cite{SSR12}, which is realized in \cite{SuSyNA}, we can show that $\mathcal{S}_*=\textrm{sup}\mathcal{C}(\zeta(L(G)),H)$ is also regular, and generated by a nondeterministic finite-state automaton $\mathcal{H}:=(Q,\Sigma_o^{\epsilon}\times\Gamma,\Xi, q_0, Q_m)$, where $Q=X\times\Sigma_o^{\epsilon}\times\Gamma\times R$ and $Q_m=X_m\times\Sigma_o^{\epsilon}\times\Gamma\times R$ with $R$ being the state set of the recognizer of $p^{-1}(p(\nu(\psi(\iota(L_{dam})))))(\Sigma_o^{\epsilon}\times\Gamma)^*$. That is $\mathcal{S}_*=L(\mathcal{H})$.   Next, we will develop a computational method to determine whether a nonblocking resilient supervisor candidate in $\mathcal{S}_*$ exists. 

To handle partial observation induced by $g$, we undertake the following subset-construction style operation on $\mathcal{H}$. Let $\mathcal{P}(\mathcal{H})=(Q_\mathcal{P},\Sigma_o^{\epsilon}\times\Gamma,\Xi_\mathcal{P},q_{0,\mathcal{P}},Q_{m,\mathcal{P}})$, where
\begin{itemize}
\item $Q_\mathcal{P}:=\Sigma_o^{\epsilon}\times 2^Q\times Q$, $Q_{m,\mathcal{P}}:=\Sigma_o^{\epsilon}\times 2^Q\times Q_m$;
\item $q_{0,\mathcal{P}}:=(\epsilon,\{q\in Q|(\exists t\in g^{-1}(\epsilon))q\in \Xi(q_0,t)\}, q_0)$;
\item The transition map $\Xi_\mathcal{P}:Q_\mathcal{P}\times\Sigma_o^{\epsilon}\times\Gamma\rightarrow 2^{Q_\mathcal{P}}$ is defined as follows: for each $(\sigma,U,q)\in Q_\mathcal{P}$ and $(\sigma',\gamma)\in \Sigma_o^{\epsilon}\times\Gamma$, if $\sigma'=\epsilon$, then 
\[\Xi_\mathcal{P}(\sigma,U,q,\epsilon,\gamma):=\{\sigma\}\times \{U\}\times \Xi(q,\epsilon,\gamma);\]
otherwise, we have
\[\Xi_\mathcal{P}(\sigma,U,q,\sigma',\gamma):=\{\sigma'\}\times \Xi(U,\sigma',\gamma)\times \Xi(q,\sigma',\gamma),\]
where \begin{center} $\Xi(U,\sigma',\gamma):=$\\ $\{\hat{q}\in Q_\mathcal{P}|(\exists \tilde{q}\in U)(\exists t\in g^{-1}(P_o(\sigma')))\hat{q}\in \Xi(\tilde{q},t)\}.$\end{center}
\end{itemize}

\textbf{Remarks:} It is clear that $L(\mathcal{P}(\mathcal{H}))=L(\mathcal{H})=\mathcal{S}_*$. In addition, since all unobservable transitions in $G_\zeta$ are selflooped at relevant states, by the construction of $\mathcal{S}_*$, we can check that the recognizer $\mathcal{H}$ also selfloops all unobservable transitions. Due to this property, we have   $\Xi_\mathcal{P}(\sigma,U,q,\epsilon,\gamma):=\{\sigma\}\times \{U\}\times\Xi(q,\epsilon,\gamma)$ in the definition of $\mathcal{P}(\mathcal{H})$, where $\Xi(q,\epsilon,\gamma)$ either equals $\{q\}$ or $\varnothing$ in $\mathcal{H}$.

To illustrate the construction procedure for $\mathcal{P}(\mathcal{H})$, assume that in Example 3 depicted in Figure \ref{fig:Cyber-Security-107}, $\mathcal{H}=G_\zeta$.
\begin{figure}[htb]
    \begin{center}
      \includegraphics[width=0.45\textwidth]{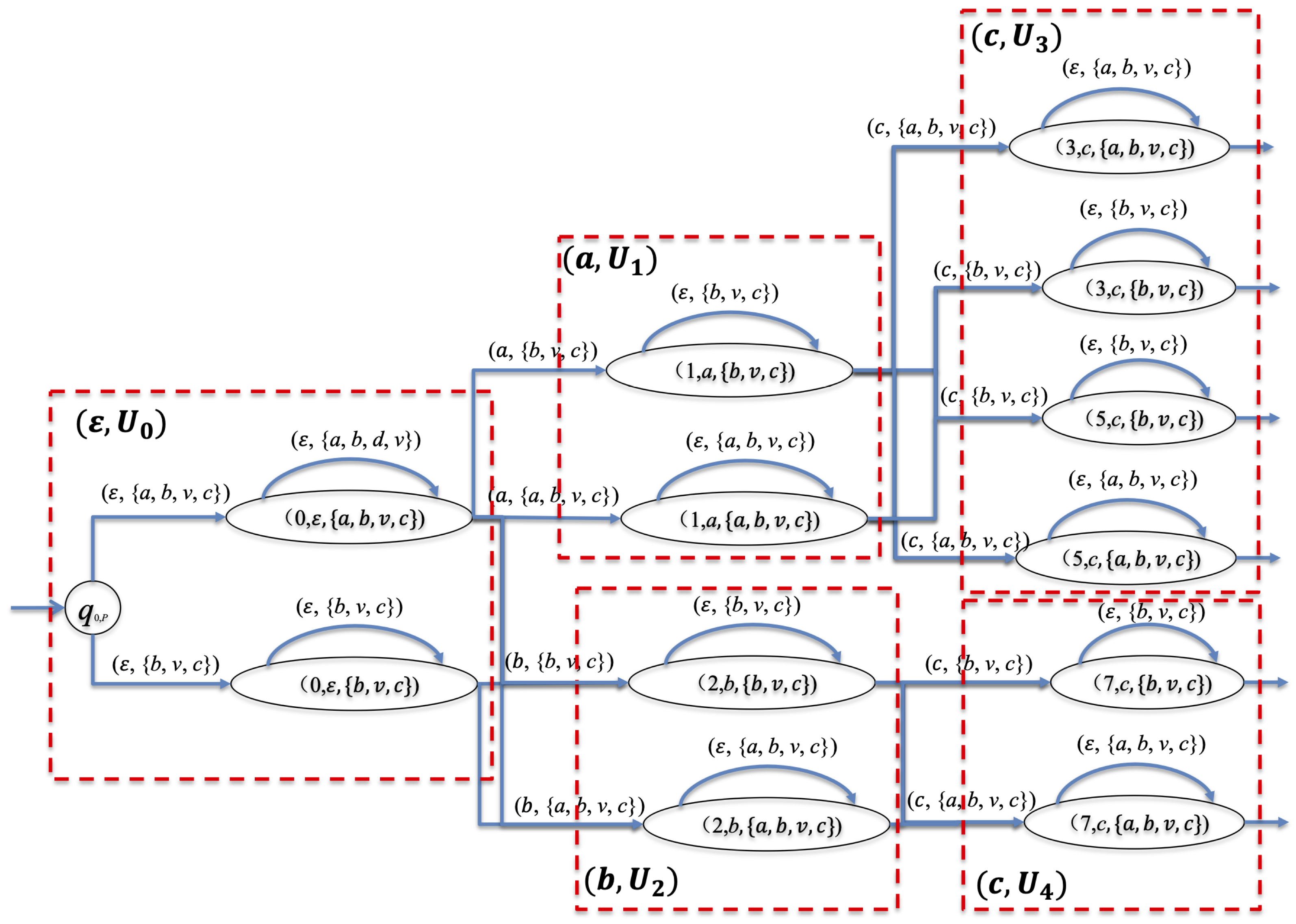}
    \end{center}
    \caption{Example 3: The model of $\mathcal{P}(\mathcal{H})$}
    \label{fig:Cyber-Security-108}
\end{figure}  
After applying the construction procedure for $\mathcal{P}(\mathcal{H})$, the outcome is depicted in Figure \ref{fig:Cyber-Security-108}, where
\begin{eqnarray*}
U_0 & = & \{q_{0,\mathcal{P}}, (0,\epsilon,\{a,b,v,c\}), (0,\epsilon,\{b,v,c\})\};\\
U_1  & = & \{(1,a,\{a,b,v,c\}), (1,a,\{b,v,c\})\};\\
U_2  & = & \{(2,b,\{a,b,v,c\}), (2,b,\{b,v,c\})\};\\
U_3  & = & \{(3,c,\{a,b,v,c\}), (3,c,\{b,v,c\}), (5,c,\{a,b,v,c\}),\\
& & (5,c,\{b,v,c\})\};\\
U_4  & = & \{(7,c,\{a,b,v,c\}), (7,c,\{b,v,c\})\}.
\end{eqnarray*}

\begin{Def}\label{Def50}\textnormal{Given $\mathcal{P}(\mathcal{H})$, a reachable sub-automaton $\Omega=(Q_\Omega\subseteq Q_\mathcal{P},\Sigma_o^{\epsilon}\times\Gamma,\Xi_\Omega,q_{0,\mathcal{P}},Q_{m,\Omega}\subseteq Q_{m,\mathcal{P}})$ of $\mathcal{P}(\mathcal{H})$ is  {\em   control feasible} if the following conditions hold:
\begin{enumerate}
\item For all $q=(\sigma,U,x,\sigma,\gamma,r)\in Q_\Omega$ with $q\neq q_{0,\mathcal{P}}$,  \[(\forall \gamma'\in\Gamma)\, \xi_\zeta(x,\sigma,\gamma,\epsilon,\gamma')\neq\varnothing \Rightarrow \Xi_\Omega(q,\epsilon,\gamma')\neq\varnothing;\]
%\xi_\zeta(x,\sigma,\gamma,\epsilon,\gamma)\neq\varnothing\Rightarrow \Xi_\Omega(q,\epsilon,\gamma)\neq\varnothing;\]
\item For all $(\sigma,U,x_1,\sigma,\gamma_1,r_1),(\sigma,U,x_2,\sigma,\gamma_2,r_2)\in Q_\Omega$, we have  $\gamma_1=\gamma_2$;
\item For each $q=(\sigma,U,x,\sigma,\gamma,r)\in Q_\Omega$,
\[g(En_\Omega(q))=P_o(\gamma)\cap g(En_{G_\zeta}(x,\sigma,\gamma));\]
\item For all $(\sigma,U,q)\in Q_\Omega$ and $\mu\in\Sigma_o^{\epsilon}\times\Gamma$, if $\Xi_\Omega(\sigma,U,q,\mu)\neq\varnothing$, then for all $(\sigma,U,q')\in Q_\mathcal{P}$,
\[\Xi_\Omega(\sigma,U,q',\mu)=\Xi_\mathcal{P}(\sigma,U,q',\mu)\subseteq Q_\Omega;\] 
%  \Xi_\Omega(q_{0,\mathcal{P}},s)=\Xi_\mathcal{P}(q_{0,\mathcal{P}},s)$;
\item $\Omega$ is co-reachable.\hfill $\Box$
\end{enumerate}}
\end{Def}
The first condition in Def. \ref{Def50} essentially states that in $\Omega$ no uncontrollable transitions allowed by $G_\zeta$ shall be disabled, which is similar to the concept of {\em state controllability} in \cite{SSR10} that handles nondeterministic transitions. Based on the construction of $\mathcal{P}(\mathcal{H})$, if $(\epsilon,\gamma')$ is allowed at state $q=(\sigma,U,x,\sigma,\gamma,r)$ in $\Omega$, then $\gamma'=\gamma$ and $\Xi_\Omega(q,\epsilon,\gamma)=\{q\}$. The second condition states that all strings observably identical in $L(\Omega)$ must result in the same control pattern. The third condition states that, for any  state in $\Omega$, an observable event is allowed at state $q$ if and only if it is allowed both by the plant $G_\zeta$ and the corresponding control pattern $\gamma$ associated with $q$. The fourth condition is similar to the concept of {\em state observability} in \cite{SSR10} to handle nondeterminism, which requires that all states in $\mathcal{P}(\mathcal{H})$ reachable by strings observably identical to some string in $L(\Omega)$, must be included in $\Omega$. The last condition is self-explained. 

\begin{figure}[htb]
    \begin{center}
      \includegraphics[width=0.45\textwidth]{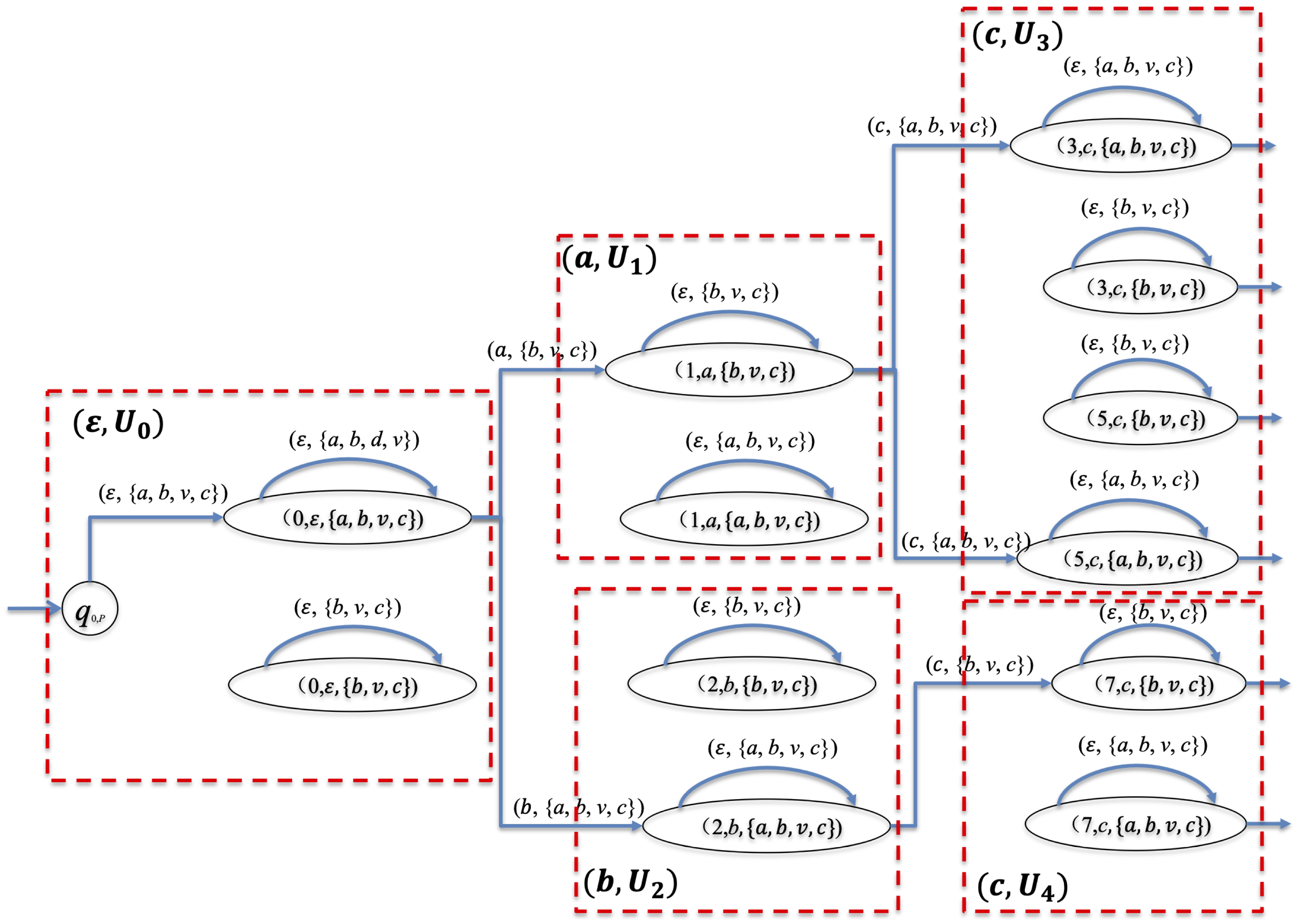}
    \end{center}
    \caption{Example 3: A model containing one $\Omega$}
    \label{fig:Cyber-Security-109}
\end{figure}

As an illustration, Figure \ref{fig:Cyber-Security-109} depicts one choice of $\Omega$ derived from $\mathcal{P}(\mathcal{H})$ in Example 3.
We can see that clearly no self-looped uncontrollable events are disabled. So the first condition in Def. \ref{Def50} holds. Due to the second condition in Def. \ref{Def50}, in $U_0$ we choose to keep $\gamma=\{a,b,v,c\}$, and thus, only states $q_{0,\mathcal{P}}$ and $(\epsilon,U_0,0,\epsilon,\{a,b,v,c\})$ will be kept in $\Omega$. Similarly, in $U_1$ the control pattern $\gamma=\{b,v,c\}$ is chosen; in $U_2$ the control pattern $\gamma=\{a,b,v,c\}$ is chosen; in $U_3$ the pattern $\gamma=\{a,b,v,c\}$ is chosen; and in $U_4$ the pattern $\gamma=\{b,v,c\}$ is chosen. Due to the third condition in Def. \ref{Def50}, we can see that in $U_0$ both outgoing transitions $(a,\{b,v,c\}$ and $(b,\{a,b,v,c\})$ of state $(\epsilon, U_0,0,\epsilon,\{a,b,v,c\})$ must be chosen in $\Omega$, as both events $a$ and $b$ are allowed by the control pattern $\{a,b,v,c\}$ and the augmented plant $G_\zeta$. In $U_1$, due to the fourth condition in Def. \ref{Def50}, both nondeterministic outgoing transitions $(c,\{a,b,v,c\})$ towards $(c,U_3,3,c,\{a,b,v,c\})$ and $(c,U_3,5,c,\{a,b,v,c\})$ must be allowed in $\Omega$. Clearly, all reachable states in $\Omega$ is co-reachable. Thus, after removing all unreachable states in Figure \ref{fig:Cyber-Security-109}, the remaining structure $\Omega$ is a control feasible sub-automaton of $\mathcal{P}(\mathcal{H})$ in Example 3. The corresponding supervisory control map $V:P_o(L(G))\rightarrow\Gamma$ can be derived as follows: $V(\epsilon):=\{a,b,v,c\}$, $V(a):=\{b,v,c\}$, $V(b):=\{a,b,v,c\}$, $V(ac):=\{a,b,v,c,\}$ and $V(bc):=\{b,v,c\}$. Similarly, we can check that in Example 2, each marked trajectory in Figure \ref{fig:Cyber-Security-106} leads to one control feasible sub-automaton $\Omega$, which satisfies all conditions in Def. \ref{Def50}.     

\begin{Theo}\label{Thm4}
\textnormal{Let $\mathcal{P}(\mathcal{H})$ be constructed as shown above. Then there exists a nonblocking resilient supervisor candidate of $\mathcal{S}_*$ if and only if there exists a control feasible reachable sub-automaton of $\mathcal{P}(\mathcal{H})$.\hfill $\Box$}
\end{Theo}

{\color{black}The complexity of computing $\mathcal{P}(\mathcal{H})$ is $O(|Q_{\mathcal{P}}|^2|\Sigma_o^\epsilon||\Gamma|)$. To determine the existence of a control feasible sub-automaton of $\mathcal{P}(\mathcal{H})$, in the worst case we need to check each sub-automaton. There are $2^{|Q_{\mathcal{P}}|}$ sub-automata. For each sub-automaton $\Omega$, whose state set is $Q_\Omega\subseteq Q_{\mathcal{P}}$, we need to check all four conditions defined in Def. \ref{Def50}, whose complexity is $O(|Q_\Omega|^2+|Q_\Omega||\Sigma_o^\epsilon||\Gamma|)$. Typically, we have $|Q_\mathcal{P}|\gg |\Sigma_o^\epsilon||\Gamma|$ and $2^{|Q_\mathcal{P}|}\gg |Q_\mathcal{P}|^3$. The final complexity of finding a control feasible sub-automaton is $O(2^{|Q_{\mathcal{P}}|})$. Notice that $|Q_\mathcal{P}|=|\Sigma_o^\epsilon|2^{|Q|}|Q|$, where $|Q|=|X||\Sigma_o^\epsilon||\Gamma||R|$ with $|R|=2^{|W|+n|\Delta_n|}$. The final complexity is $O(2^{|\Sigma_o^\epsilon|2^{|X||\Sigma_o^\epsilon||\Gamma|2^{|W|+n|\Delta_n|}}|X||\Sigma_o^\epsilon||\Gamma|2^{|W|+n|\Delta_n|}})$.} 

\begin{Theo}\label{Thm5}
\textnormal{Given a plant $G$ and a damage language $L_{dam}\subseteq L(G)$, it is decidable whether there exists a nonblocking supervisor $V$ such that the closed-loop system $V/G$ is not attackable with respect to $L_{dam}$.\hfill $\Box$}
\end{Theo}
Proof: By Theorem \ref{Thm3}, there exists a nonblocking supervisor which disallows any regular smart sensor attack with respect to $L_{dam}$ if and  only if there exists a nonblocking resilient supervisor candidate $\mathcal{L}\subseteq\mathcal{S}_*$. By Theorem \ref{Thm4}, we know that there exists a nonblocking resilient supervisor candidate if and only if there exists a control feasible sub-automaton of $\mathcal{P}(\mathcal{H})$, which recognizes $\mathcal{S}_*$. Since there exists a finite number of sub-automata in  $\mathcal{P}(\mathcal{H})$, the existence of a control feasible sub-automaton of $\mathcal{P}(\mathcal{H})$ is decidable. Thus, the existence of a nonblocking supervisor which disallows any regular smart sensor attack with respect to $L_{dam}$  is decidable. \hfill $\blacksquare$

It is interesting to point out that, in general, there are typically many choices of a control feasible sub-automaton $\Omega$, leading to possibly many resilient supervisors. It is unfortunate that the most permissive resilient supervisor in terms of set inclusion of closed-loop behaviours typically does not exist. For example, in Example 2 there are up to three different supervisory control maps depicted in Figure \ref{fig:Cyber-Security-106}, leading to two non-compatible maximally permissive supervisors: one generates the closed-loop behaviour of $L(V_1/G)=\overline{\{avc\}}$ and the other one generates $L(V_2/G)=\overline{\{bvd\}}$. It is an interesting question whether the structure $\mathcal{P}(\mathcal{H})$ could be used to directly synthesize a maximally permissive nonblocking resilient supervisor, as it conceptually contains all resilient supervisors.

\section{Conclusions}
Although in our early work \cite{Su2017} \cite{Su2018}, the concept of smart sensor attacks was introduced, and syntheses of a smart sensor attack and a supervisor resilient to smart sensor attacks were presented, it has not been shown whether the existence of a nonblocking supervisor resilient to {\color{black}all} smart sensor attacks is decidable, as the synthesis algorithm presented in \cite{Su2018} does not guarantee to find a resilient supervisor, even though it may exist. In this paper we have first shown that the existence of a regular smart weak sensor attack is decidable, and in case it exists, it can be synthesized. Our first contribution is to  identify risky pairs that describe how a legal sequence of control patterns may be used by a sensor attack to inflict weak damage, which is stated in Theorem 1 that there exists a smart weak sensor attack if and only if there exists at least one risky pair. Notice that this result is valid, regardless of whether the attack model is regular, i.e., representable by a finite-state automaton. With this key idea, to ensure the existence of a supervisor resilient to smart sensor attacks, we only need to make sure that there should be no risky pairs. Our second contribution is to show that all risky pairs can be identified and removed from the plant behaviours, via a genuine encoding scheme, upon which a verifiable sufficient and necessary condition is presented to ensure the existence of a nonblocking supervisor resilient to smart sensor attacks. This establishes the result that the existence of a supervisor resilient to {\color{black}all} smart sensor attacks is decidable. Finally, as ou{\color{black}r} third contribution, the decision process renders a synthesis algorithm for a resilient supervisor, whenever it exists, which has never been addressed in any existing works. 

The decidability result established in this paper may shed light on future research on cyber attack related resilient synthesis, e.g., to decide existence of a resilient supervisor for smart actuator attacks or smart attacks with observations different from those of the supervisor, which are gaining more and more attention recently. This decidability result allows us to focus more on computational efficiency related to smart sensor attacks. \\           
     
\noindent \textbf{Appendix}\\

\noindent 1. Proof of Theorem 1: (1) We first show the IF part. Assume that there exist $s=u_1\sigma_1\cdots u_r\sigma_r u_{r+1}\in L_{dam}$, with $r\in\mathbb{N}$, $u_1,\cdots, u_{r+1}\in\Sigma_{uo}^*$ and $\sigma_1,\cdots,\sigma_r\in\Sigma_o$, and $t=\nu_1\cdots\nu_r\in P_o(L(V/G))$ with $\nu_1,\cdots,\nu_r\in\Delta_n$ such that (1) $u_1,\sigma_1\in V(t^0)^*$; (2) for each $i\in \{2,\cdots,r\}$, $u_i,\sigma_i\in V(t^{i-1})^*$; (3) $u_{r+1}\in V(t)^*$. We now explicitly design an attack model $A$ as follows.
\begin{enumerate}
\item $A(\epsilon):=\epsilon$;
\item for each $s\in P_o(L(G))$, where $A(s)$ has been defined, for each $\sigma\in\Sigma_o$ with $s\sigma\in P_o(L(G))$,
\[A(s\sigma):=\left\{\begin{array}{ll} A(s)\nu_i & \textrm{if $s\sigma=\sigma_1\cdots\sigma_i$, $i\in\{1,\cdots,r\}$;}\\
A(s)\sigma & \textrm{if $s\sigma\in L(V/G)-\overline{\{\sigma_1\cdots\sigma_r\}};$}\\
A(s) & \textrm{otherwise.}\end{array}\right.\]
\end{enumerate}
{\color{black}Clearly, $A$ is well defined.} We now show that $A(P_o(L(V\circ A/G)))\subseteq L(S)$, i.e., $A$ is covert, by using induction on the length of strings in $P_o(L(G))$. Clearly, $\epsilon\in P_o(L(V\circ A/G))$, and $A(\epsilon)=\epsilon\in L(S)$. Assume that for all strings $s\in P_o(L(V\circ A/G))$ with $|s|\leq n$, where $n\in\mathbb{N}$, we have $A(s)\in L(S)$. We need to show that for each $\sigma\in\Sigma_o$ with $s\sigma\in P_o(L(V\circ A/G))$, we have $A(s\sigma)\in L(S)$. If $s\sigma=\sigma_1\cdots\sigma_{n+1}$, where $n+1\leq r$, then we have $A(s\sigma)=\nu_1\cdots\nu_{n+1}\in \overline{\{t\}}\subseteq P_o(L(V/G))\subseteq L(S)$. If $s\sigma\in L(V/G)-\overline{\{\sigma_1\cdots\sigma_r\}}$, {\color{black}then we have two cases to consider. Case 1:} $A(s)\sigma\in L(V/G)\subseteq L(S)$. Then $A(s\sigma)=A(s)\sigma\in L(S)$.  {\color{black}Case 2:} $A(s)\sigma\notin L(S)$. We will show that, $s\sigma\notin P_o(L(V\circ A/G))$. Since {\color{black}$s\sigma\in L(V\circ A/G)$, we have}  $A(s)\subseteq L(S)$. Because $s\in P_o(L(G))$, there must exist $\tilde{s}\in L(V\circ A/G)$ such that $s=P_o(\tilde{s})$. For all $\tilde{u}\in\Sigma_{uo}^*$, if $\tilde{s}\tilde{u}\sigma\in L(V\circ A/G)$, we know that $\tilde{u}\sigma\in V(A(P_o(\tilde{s})))^*=V(A(s))^*$. By the definition of $V$, we know that $\sigma$ must also be in $V(A(s))$. Thus, $A(s)\sigma\in L(S)$. But this contradicts our assumption that $A(s)\sigma\notin L(S)$. Thus, the only possibility is that $s\sigma\notin {\color{black}L(V/G)}$. {\color{black}Since $s\sigma\in P_o(L(V\circ A/G))$, we know that $A(s)\in L(S)$ and $\sigma\in V(A(s))\in\Gamma(V)$. If $s\in L(V/G)$, then clearly $s\sigma\in L(G)-(L(V/G)\cup \overline{\{\sigma_1\cdots\sigma_r\}})$. Thus, by Definition of $A$, we have $A(s\sigma)=A(s)\in L(S)$. With a similar argument, we know that for all $s'\in\Sigma^*$, $ss'\in L(V\circ A/G)$ implies that $s'\in V(A(s))^*$, and $A(ss')=A(s)\in L(S)$. If $s\notin L(V/G)$, then we can always find $\hat{s}\leq s$ and $\hat{\sigma}\in\Sigma_o$ with $\hat{s}\hat{\sigma}\leq s$ such that $\hat{s}\in L(V/G)$ but $\hat{s}\hat{\sigma}\notin L(V/G)$. Then with the same argument, we know that $(s/\hat{s})\sigma\in L(G)-(L(V/G)\cup \overline{\{\sigma_1\cdots\sigma_r\}})$, namely $A(s\sigma)=A(\hat{s}(s/\hat{s})\sigma)=A(\hat{s})\in L(S)$. Thus,  the induction part holds, which means $A$ is covert.}   

Since $A$ results in weak damage due to the existence of $\hat{s}$, by Def. \ref{Def20}, $A$ is a smart weak sensor attack.\\    
(2) Next, we show the ONLY IF part. Assume that there exists a smart weak sensor attack $A$. By Def. 1, we know that $A(P_o(L(V\circ A/G)))\subseteq L(S)$ and $L(V\circ A/G)\cap L_{dam}\neq\varnothing$. Thus, there exists $s=u_1\sigma_1\cdots u_r\sigma_r u_{r+1}\in L(V\circ A/G)\cap L_{dam}$ with $r\in\mathbb{N}$, $u_1,\cdots, u_{r+1}\in\Sigma_{uo}^*$ and $\sigma_1,\cdots,\sigma_r\in\Sigma_o$, such that $A(P_o(u_1))=\epsilon$, $A(P_o(u_1)\sigma_1)=\nu_1\in\Delta_n$; $A(P_o(u_1)\sigma_1\cdots P_o(u_j))=A(P_o(u_1)\sigma_1\cdots P_o(u_{j-1})\sigma_{j-1})$ and $A(P_o(u_1)\sigma_1\cdots P_o(u_j)\sigma_j)=\nu_1\cdots\nu_j$ with $\nu_j\in\Delta_n$ for all $j\in\{2,\cdots, r\}$; and finally, \[A(P_o(s))=A(P_o(u_1)\sigma_1\cdots P_o(u_r)\sigma_r).\] Let $t=\nu_1\cdots \nu_r$. Since $s\in L(V\circ A/G)$, by the definition of $L(V\circ A/G)$, we know that  (1) $u_1,\sigma_1\in V(t^0)^*$; (2) for each $i\in \{2,\cdots,r\}$, $u_i,\sigma_i\in V(t^{i-1})^*$; (3) $u_{r+1}\in V(t)^*$.  Thus, the theorem follows. \hfill $\blacksquare$ 

\noindent 2. Proof of theorem 2: Since $V$ is regular, there is a finite-state automaton $S=(Z,\Sigma,\delta,z_0,Z_m=Z)$ that realizes $V$. We follow an idea adopted from \cite{Su2018}, and start with a single-state transducer $\mathcal{A}=(Y,{\color{black}\Sigma_o^{\epsilon}}\times\Delta_n,\eta,I,O,y_0,Y_m=Y)$, where $Y=\{y_0\}$ and for all $(\sigma,u)\in\Sigma_o\times\Delta_n$, we have $\eta(y_0,\sigma,u)=y_0$. $\mathcal{A}$ contains all possible attack moves. Since $L_{dam}$ is regular, there exists a finite-state automaton $D=(W,\Sigma,\kappa,w_0,W_m)$ such that $L(D)=\Sigma^*$ and $L_m(D)=L_{dam}$. We now form a combination of all relevant finite-state transition structures. Let
\[\Psi=(N, \Sigma_N, \lambda, n_0, N_m),\]       
where
\begin{itemize}
\item $N=X\times Z\times Y\times W$, $N_m=X_m\times Z_m\times Y_m\times W_m$;
\item $n_0=(x_0,z_0,y_0,w_0)$;
\item $\Sigma_N:=\{(\sigma,\sigma',u)\in\Sigma\times \Sigma_o^{\epsilon}\times\Delta_n|\sigma'=P_o(\sigma)\wedge [\sigma'=\epsilon\Rightarrow u=\epsilon]\}$;
\item $\lambda:N\times\Sigma_N\rightarrow N$ is the partial transition map: for each $(x,z,y,w), (x',z',y',w')\in N$ and $(\sigma,\sigma',u)\in \Sigma_N$, 
\[\lambda(x,z,y,w,\sigma,\sigma',u)=(x',z',y',w')\]
if and only if the following conditions hold:
\begin{enumerate}
\item $\sigma\in En_S(z)$;
\item $\delta(z,u)!$;
\item $\xi(x,\sigma)=x'$, $\delta(z,u)=z'$, $\eta(y,\sigma',u)=y'$, and $\kappa(w,\sigma)=w'$.
\end{enumerate}
\end{itemize}
Let the controllable alphabet  of $\Psi$ be \[\Sigma_{N,c}:=\Sigma_N\cap (\Sigma_o\times\Sigma_o\times\Delta_n).\] Let $\pi:\Sigma_N^*\rightarrow\Sigma^*$ be a projection, where
\begin{itemize}
\item $\pi(\epsilon)=\epsilon$;
\item $(\forall (\sigma,\sigma',u)\in\Sigma_N)\, \pi(\sigma,\sigma',u):=\sigma$;
\item $(\forall s\varsigma\in\Sigma_N^*)\, \pi(s\varsigma)=\pi(s)\pi(\varsigma)$.
\end{itemize}
Similarly, let $\varpi:\Sigma_N^*\rightarrow\Delta_n^*$ be a projection, where
\begin{itemize}
\item $\varpi(\epsilon)=\epsilon$;
\item $(\forall (\sigma,\sigma',u)\in\Sigma_N)\, \varpi(\sigma,\sigma',u):=u$;
\item $(\forall s\varsigma\in\Sigma_N^*)\, \varpi(s\varsigma)=\varpi(s)\varpi(\varsigma)$.
\end{itemize}
We calculate a controllable \cite{RW87} prefix-closed sublanguage $U\subseteq L(\Psi)$ w.r.t. $\Psi$ and $\Sigma_{N,uc}:=\Sigma_N-\Sigma_{N,c}$, i.e.,
\[U\Sigma_{N,uc}\cap L(\Psi)\subseteq U,\]
which satisfies the following properties:
\begin{itemize}
\item $\pi(U)\cap L_m(D)\neq\varnothing$;
%\item $\pi(U)$ is observable with respect to $L(G)$ and $P_o$;
\item $(\forall s\in U)\pi(\{\varsigma\in \Sigma_N|s\varsigma\in U\})= En_{L(G)}(\pi(s))\cap En_S(\varpi(s))$;
\item $(\forall s\in U)(\forall t\in \pi^{-1}(P_o^{-1}(P_o(\pi(s))))\cap U)\varpi(s)=\varpi(t)$. 
\end{itemize}

We can check that Condition 1 of $U$ states a weak nonblocking property. Condition 2 is an ``extended'' controllability property, which states that, after a string $s\in U$, each outgoing transition $\varsigma\in \Sigma_N$ is allowed, as long as the plant $G$ allows it, i.e., $\pi(\varsigma)\in En_{L(G)}(\pi(s))$, and the supervisor also allows it, i.e., $\pi(\varsigma)\in En_S(\varpi(s))$. Because the supervisor allows all uncontrollable transitions, thus, no uncontrollable events in $\Sigma_{uc}$ shall be disabled here, which is the reason why we call it a special controllability property. Condition 3 states that any two strings $s$ and $t$, ``observably'' identical in the sense that $P_o(\pi(s))=P_o(\pi(t))$, must lead to the same (fake) observable strings in $\Delta_n^*$, i.e., $\varpi(s)=\varpi(t)$, which means $P_o(\pi(s))=P_o(\pi(t))$ implies the same attack move - thus, may result in a deterministic attack function. Based on this interpretation, by adopting either a power set construction over $\Psi$ via the projection $P_o\circ\pi$ and a state pruning algorithm similar to the one proposed in \cite{SSR10} that originally aims to compute a supremal nonblocking supervisor with respect to $L_m(D)$, that is state-controllable and state-normal, or one algorithm proposed in \cite{YL16} that computes a maximally controllable and observable nonblocking supervisor,  we can show that the existence of such a $U$ is decidable. 

To complete the proof, we only need to show that there exists a regular smart weak sensor attack if and only if there exists such a language $U$. 

To show the IF part, we define an attack model $A:P_o(L(G))\rightarrow \Delta_n^*$, where 
\begin{itemize}
\item $(\forall s\in U)A(P_o(\pi(s))):=\varpi(s)$;
\item for all $s\in U$ and $\mu \in\Sigma_N$, \[s\mu\notin U\Rightarrow (\forall t\in\Sigma_N^*)A(P_o(\pi(s\mu t))):=A(P_o(\pi(s))).\]
\end{itemize}
Since for all $s,t\in U$, if $P_o(\pi(s))=P_o(\pi(t))$, then $\varpi(s)=\varpi(t)$, we know that $A$ is well defined. 

Next, we show that conditions (1) {\color{black}and} (3) stated in Def. \ref{Def20} hold for $A$, meaning that $A$ is a smart weak sensor attack.   

We first show that Condition (1) in Def. \ref{Def20} holds, i.e., $A(P_o(L(V\circ A/G)))\subseteq L(S)$. We first consider all $s\in U$.  Since $U\subseteq L(\Psi)$,  by the definition of $\Psi$ and Condition 2 of $U$, we write string $s$ as $s={\color{black}s_1\cdots s_r\in U}$, where {\color{black}$s_i=(\tau_{i,1},\epsilon,\epsilon)\cdots (\tau_{i,l_i},\epsilon,\epsilon)(\sigma_i,\sigma_i,u_i)$ with $r, l_1,\cdots,l_r\in\mathbb{N}$, $1\leq i\leq r$, $\sigma_1,\cdots,\sigma_r\in\Sigma_o$ and $\{\tau_{i,1},\cdots,\tau_{i,l_i}\}\subseteq\Sigma_{uo}$ such that $\tau_{j,p}\in V(u_1\cdots u_{j-1})=En_S(\delta(z_0,u_1\cdots u_{j-1}))$ ($j=2,\cdots, r$, $1\leq p\leq l_j$) and $\tau_{1,q}\in V(\epsilon)=En_S(z_0)$ with $1\leq q\leq l_1$.} Thus,  $\varpi(s)=u_1\cdots u_r\in L(S)$, as all unobservable transitions are self-looped in $S$. Thus, $A(P_o(\pi(s)))=\varpi(s)\in L(S)$. For each string $t\in L(V\circ A/G)$, there exists $s'\in \pi^{-1}(t)\in\Sigma_N^*$ and, by the definition of $A$, $A(P_o(\pi(s')))=A(P_o(t))\subseteq L(S)$. 

Next, we show that Condition (3) in Def. \ref{Def20} holds, that is,  $L(V\circ A/G)\cap L_{dam}\neq\varnothing$. To see this, notice that $L(V\circ A/G)=\pi(U)$ and since  $\pi(U)\cap L_{dam}\neq\varnothing$, we have $L(V\circ A/G)\cap L_{dam}\neq\varnothing$. {\color{black}Thus, by} Def. \ref{Def20}, $A$ is a smart weak sensor attack.

To show the ONLY IF part, assume that there exists a regular smart weak sensor attack $A$ represented by a finite-transducer $\mathcal{A}=(Y,\Sigma_o^{\epsilon}\times\Delta_n,\eta,I,O,y_0,Y)$. We construct $\Psi$ as shown above and let $U=L(\Psi)$. Since $\mathcal{A}$ is a smart weak sensor attack, conditions (1) {\color{black}and} (3)  in Def. \ref{Def20} hold. Clearly, $U$ is controllable and prefix closed. We know that $\pi(U)\cap L_{dam}=L(V\circ A/G)\cap L_{dam}\neq\varnothing$, due to Condition (3) in Def. \ref{Def20}. Due to the covertness condition in Def. \ref{Def20}, we know that, for all $s\in U$, we have $\varpi(s)\in L(S)$. By the definition of $U$, we know that
\[\pi(\{\varsigma\in \Sigma_N|s\varsigma\in U\})= En_{L(G)}(\pi(s))\cap En_S(\varpi(s)).\]
Finally, because the attack model $A:P_o(L(G))\rightarrow \Delta_n^*$ is a map, which maps all strings observably identical to the same observable string acceptable by $S$, by the definition of $U$, we have 
\[(\forall s\in U)(\forall t\in \pi^{-1}(P_o^{-1}(P_o(\pi(s))))\cap U)\varpi(s)=\varpi(t).\]
Thus, all required conditions for $U$ hold. This completes the proof. \hfill $\blacksquare$ 

\noindent 3. Proof of Theorem 3: (1) We first show the IF part. Assume that there exists a nonblocking resilient supervisor candidate $\mathcal{L}\subseteq \mathcal{S}_*$. For each $s\in P_o(L(G))$, if $s\notin g(\mathcal{L})$ then let $V(s):=\Sigma_{uc}$; otherwise, for any $u\in g^{-1}(s)\cap \mathcal{L}$, let  $V(s):=[p(u)]^{\uparrow}$. For the latter case, we first show that $V(s)$ is well defined. Assume that it is not true, then there exist $u_1,u_2\in g^{-1}(s)\cap\mathcal{L}$ such that $u_1\neq u_2$ and $[p(u_1)]^{\uparrow}\neq [p(u_2)]^{\uparrow}$. But this violates Condition 1 of Def. \ref{Def4}, thus, contradicts our assumption that  $\mathcal{L}$ is a nonblocking resilient supervisor candidate. So $V$ must be well defined, that is, for each $s\in P_o(L(G))$, $V(s)$ is uniquely defined.    

Secondly, since $[p(u)]^{\uparrow}$ is a control pattern for $u\in g^{-1}(s)\cap\mathcal{L}$, it is clear that $V(s)\in\Gamma$. Since $V$ maps all strings observably identical to a same control pattern, we know that $L(V/G)$ is observable. Finally, by the third condition of Def. \ref{Def4}, it is clear that $\mathcal{L}$ is nonblocking. By the construction of $\mathcal{L}$, {\color{black}and the second condition of Def. \ref{Def4}, where for all $\hat{s}\in\mathcal{L}$, 
\[g(En_{\mathcal{L}}(\hat{s}))= P_o(p(\hat{s}^{\uparrow}))\cap g(En_{\zeta(L(G))}(\hat{s})),\]
we can show that $g(\mathcal{L})= P_o(L(V/G))$.}  Thus, by the third condition of Def. \ref{Def4}, we have  that $V$ is a nonblocking supervisory control map. Clearly, $V$ does not allow any weak sensor attack damage. Thus, it is resilient to any smart sensor attack, regardless of whether the attack is a strong or weak one. 

\noindent (2) We now show the ONLY IF part. Assume that there exists a supervisor $V$, which does not allow any smart sensor attack. Since each strong attack is also a weak attack, we will only need to consider weak sensor attacks. We define the following language $\mathcal{L}$ induced from $V$: 
\begin{enumerate}
\item[i)] $\epsilon\in\mathcal{L}$;
\item[ii)] $(\epsilon, V(\epsilon))\in\mathcal{L}$;
\item[iii)] For all $s\in\mathcal{L}$, 
%         \begin{itemize}
%         \item $p(s)^{\uparrow}\cap\Sigma_{uo}\neq\varnothing\iff \{s\}\{(\epsilon,p(s)^{\uparrow})\}^*\subseteq\mathcal{L}$;
and $\sigma'\in P_o(p(s)^{\uparrow})$ and $\gamma\in\Gamma$, 
         \[(\sigma',\gamma)\in En_{\zeta(L(G))}(s)\Rightarrow s(\sigma',V(g(s)\sigma'))\in\mathcal{L};\]
%         \end{itemize}
\item[iv)] All strings in $\mathcal{L}$ are generated in Steps (1)-(3).
\end{enumerate}
Clearly, $\mathcal{L}\subseteq\zeta(L(G))$. Because $V$ is a resilient supervisor, by Theorem \ref{thm0} we know that $\mathcal{L}\subseteq H$ - otherwise, there must exist a smart weak attack. By the construction of $\mathcal{L}$, we know that  $\mathcal{L}$ is conditionally controllable  with respect to $\zeta(L(G))$ and $\{\epsilon\}\times\Gamma$. Thus, $\mathcal{L}\in\mathcal{C}(\zeta(L(G)),H)$, namely, $\mathcal{L}\subseteq \mathcal{S}_*$. Since $V$ is a nonblocking supervisor, we can check that {\color{black}the first and last conditions} in Def. \ref{Def4} hold. {\color{black}Since $\mathcal{L}\subseteq\zeta(L(G))$, we know that $g(En_{\mathcal{L}}(s))\subseteq g(En_{\zeta(L(G))}(s))$. By Steps (iii)-(iv), we know that $g(En_{\mathcal{L}}(s))\subseteq P_o(p(s^{\uparrow}))$. Thus, we have $g(En_{\mathcal{L}}(s))\subseteq P_o(p(s^{\uparrow}))\cap g(En_{\zeta(L(G))}(s))$. On the other hand, by Step (iii), we know that  $g(En_{\mathcal{L}}(s))\supseteq P_o(p(s^{\uparrow}))\cap g(En_{\zeta(L(G))}(s))$. Thus, we finally have
\[g(En_{\mathcal{L}}(s))= P_o(p(s^{\uparrow}))\cap g(En_{\zeta(L(G))}(s)),\]
which means the second condition of Def. \ref{Def4} holds.} Thus, $\mathcal{L}$ is a nonblocking resilient supervisor candidate, which completes the proof.\hfill $\blacksquare$

\noindent 4. Proof of Theorem 4: (1) To show the IF part, assume that $\Omega$ is a control feasible reachable sub-automaton of $\mathcal{P}(\mathcal{H})$. Let $\mathcal{L}:=L(\Omega)$. By condition (1) of Def. \ref{Def50}, we know that $\mathcal{L}$ is conditionally controllable with respect to $\zeta(L(G))$ and $\{\epsilon\}\times\Gamma$. Thus, $\mathcal{L}\in\mathcal{C}(\zeta(L(G)),H)$. For all $s\in \mathcal{L}$ and $t\in g^{-1}(g(s))\cap\mathcal{L}$, let $(\sigma,U,x_1,\sigma,\gamma_1,r_1), (\sigma,U,x_2,\sigma,\gamma_2,r_2)\in Q_\Omega$ be induced by $s$ and $t$ with $\sigma=g(s)^{\uparrow}=g(t)^{\uparrow}$. Then by condition (2) of Def. \ref{Def50},  we have $\gamma_1=\gamma_2$. Thus, the first condition in Def. \ref{Def4} holds. In addition, we have
\[En_{\mathcal{L}}(s):=\bigcup_{q=(g(s)^{\uparrow},U,x,g(s)^{\uparrow},\gamma,r)\in \Xi_\Omega(q_{0,\mathcal{P}},s)}En_\Omega(q).\]
By condition (3) of Def. \ref{Def50}, we have $g(En_\Omega(q))=P_o(\gamma)\cap g(En_{G_\zeta}(x,g(s)^{\uparrow},\gamma))$,
by condition (4) of Def. \ref{Def50}, we have
\[\bigcup_{(\sigma,U,x,\sigma,\gamma,r)\in \Xi_\Omega(q_{0,\mathcal{P}},s)}En_{G_\zeta}(x,\sigma,\gamma)=En_{\zeta(L(G))}(s),\]
where $\sigma=g(s)^{\uparrow}$. Thus, we conclude that
$g(En_{\mathcal{L}}(s))= P_o(p(s^{\uparrow}))\cap g(En_{\zeta(L(G))}(s))$, namely the second condition of Def. \ref{Def4} holds.  Finally, since $\Omega$ is co-reachable, and together with condition (4) of Def. \ref{Def50}, we know that $\mathcal{L}$ is nonblocking. Thus, by Def. \ref{Def4}, $\mathcal{L}$ is a nonblocking resilient supervisor candidate of $\mathcal{S}_*$.  

\noindent (2) To show the ONLY IF part, assume that there exists a nonblocking resilient supervisor candidate $\mathcal{L}\subseteq\mathcal{S}_*$. We need to show that there exists a control feasible sub-automaton $\Omega$ of $\mathcal{P}(\mathcal{H})$. 
We construct a sub-automaton $\mathcal{P}(\mathcal{H})_\mathcal{L}:=(Q_\mathcal{L},\Sigma_o^{\epsilon}\times\Gamma,\Xi_\mathcal{L},q_{0,\mathcal{P}},Q_{m,\mathcal{L}})$, where 
\[Q_{\mathcal{L}}:=\{q\in Q_\mathcal{P}|(\exists s\in \mathcal{L}) q\in \Xi_\mathcal{P}(q_{0,\mathcal{P}},s)\},\]
and $Q_{m,\mathcal{L}}:=Q_\mathcal{L}\cap Q_{m,\mathcal{P}}$. The transition map $\Xi_\mathcal{L}$ is the restriction of $\Xi_\mathcal{P}$ over $Q_\mathcal{L}$.

Let $\Omega$ be the sub-automaton $\mathcal{P}(\mathcal{H})_\mathcal{L}$. Since $\mathcal{L}$ is a supervisor candidate, by the first condition of Def. \ref{Def4}, we have the following property: 
\[(\forall s\in\mathcal{L})\{[p(t)]^{\uparrow}|t\in g^{-1}(g(s))\cap \mathcal{L}\}=\{p(s)^{\uparrow}\}. \hspace{1cm} (*)\]
By the construction of $\mathcal{P}(\mathcal{H})$, we know that for each state reachable by $s$, say $(g(s)^{\uparrow},U_s,q_s)$, and each state reachable by $t\in g^{-1}(g(s))\cap\mathcal{L}$, say $(g(t)^{\uparrow},U_t,q_t)$, we have $U_s=U_t$. Thus, if $q_s=(x_s,g(s)^{\uparrow},\gamma_s,r_s)$ and $q_t=(x_t,g(t)^{\uparrow},\gamma_t,r_t)$, by the property $(*)$, we have $\gamma_s=\gamma_t$, which means the second condition of Def. \ref{Def50} holds. Based on the construction of $\Omega$,  it is also clear that the condition (1) of Def. \ref{Def50} holds because $\mathcal{P}(\mathcal{L})_{\mathcal{L}}$ is conditionally controllable due to the conditional controllability of  $\mathcal{L}$. Because $\mathcal{P}(\mathcal{H})_\mathcal{L}$ is derived from a language $\mathcal{L}$, the fourth condition of Def. \ref{Def50} holds for $\mathcal{P}(\mathcal{H})_\mathcal{L}$. In addition, since $\mathcal{L}$ is a resilient supervisor candidate, by the second condition of Def. \ref{Def4}, we know that the third condition of Def. \ref{Def50} holds. Finally, since $\mathcal{L}$ is nonblocking, based on Def. \ref{Def5}, we know that each state in $\Omega$ must be co-reachable. This completes the proof that $\Omega$ is indeed control feasible. \hfill $\blacksquare$

\end{document}